\documentclass[12pt,preprint]{aastex}

\slugcomment{The Astronomical Journal, in press 2009-Feb-24}

\shorttitle{Active Centaurs}
\shortauthors{Jewitt}

\received{not yet}
\begin{document}

\title{The Active Centaurs}
\shortauthors{Jewitt}


\author{David Jewitt}
\affil{Institute for Astronomy, University of Hawaii, \\
2680 Woodlawn Drive, Honolulu, HI 96822}

\email{jewitt@hawaii.edu}

\begin{abstract}
The Centaurs are recent escapees from the Kuiper belt that are destined either to meet fiery oblivion in the hot inner regions of the Solar system or to be ejected to the interstellar medium by gravitational scattering from the giant planets.  Dynamically evolved Centaurs, when captured by Jupiter and close enough to the Sun for near-surface water ice to sublimate, are conventionally labeled as ``short-period'' (specifically, Jupiter-family) comets.
Remarkably, some Centaurs show comet-like activity even when far beyond the orbit of Jupiter, suggesting mass-loss driven by a process other than the sublimation of water ice.  We observed a sample of  23 Centaurs and found nine to be active, with mass-loss rates measured from several kg s$^{-1}$ to several tonnes s$^{-1}$.  Considered as a group, we find that the 
``active Centaurs'' in our sample have perihelia smaller than the inactive Centaurs (median 5.9 AU vs.\ 8.7 AU), and smaller than the median perihelion distance computed for all known Centaurs (12.4 AU).  This suggests that their activity is thermally driven.  We consider several possibilities for the origin of the mass-loss from the active Centaurs.   Most are too cold for activity at the observed levels to originate via the sublimation of crystalline water ice.  Solid carbon monoxide and carbon dioxide have the opposite problem: they are so volatile that they should drive activity
in Centaurs at much larger distances than observed.  We consider the possibility that activity in the Centaurs is triggered by the conversion of amorphous ice into the crystalline form accompanied by the release of trapped gases, including carbon monoxide.  By imposing the condition that crystallization should occur when the crystallization time is shorter than the orbital period we find a qualitative match to the perihelion distribution of the active Centaurs and conclude that the data are consistent with the hypothesis that the Centaurs contain amorphous ice.  

\end{abstract}


\section{Introduction}
The Centaurs experience repeated close encounters with the giant planets and, as a result, follow orbits which are both chaotic and dynamically short-lived compared to the age of the Solar system.  Numerical simulations give half-lives from 0.5 Myr to 30 Myr, with average values increasing with semimajor axis and perihelion distance (Horner et al.\ 2004) and a population median lifetime near 10 Myr (Tiscareno and Malhotra 2003). 
The short dynamical lifetimes suggest that the Centaurs must be resupplied from a more stable source region elsewhere, if their population
is to remain in steady-state.  This source region is most likely the Kuiper belt, from which objects drift first into Neptune-crossing trajectories and then are scattered between the giant planets (see Morbidelli 2008 for a recent review of Kuiper belt dynamics).   The Kuiper belt contains several dynamically distinct classes of body, and it is not clear precisely from where in the belt the Centaurs are derived.  The source is popularly identified with the so-called scattered Kuiper belt objects but other trans-Neptunian populations, or even completely different source regions, may be implicated  (c.f.\  Volk and Malhotra 2008).  The Centaurs suffer several fates.  A few percent of the Centaurs impact the giant planets, or are
broken apart by gravitational stresses exerted upon them by near-miss interactions with the giants (the famous comet
D/Shoemaker-Levy 9 is the best example of a body suffering both of these fates).   About two thirds are ultimately ejected from the Solar system into interstellar space (Tiscareno and Malhotra 2003).  The remaining one third become temporarily trapped in small, near-Sun orbits, where they sublimate and are conventionally labeled as ``Jupiter family comets''.  The Centaurs, then,
are scientifically interesting as a kind of intermediate population, halfway between the distant, frigid and relatively stable bodies
in the Kuiper belt and the rapidly sublimating, disintegrating comets of the hot inner regions (see Jewitt 2004 for a broad account of the interrelations between these objects).  

Surprisingly, there is no uniformly accepted definition of the Centaurs.    Some dynamicists classify all objects with perihelia between the orbits of Jupiter and Neptune as Centaurs, regardless of their semimajor axes.  This definition is reasonable, given that the dynamics are dominated by perihelic interactions with the planets but some qualifications are needed, since many resonant Kuiper belt objects (like 134340 Pluto), have perihelia inside Neptune's orbit but are not usefully considered as Centaurs.   A detailed classification system has been proposed by Horner et al.\ (2004) in which the Centaurs are labeled by the planets with which they most strongly interact.   We
here adopt a simpler operational definition, modified from Jewitt and Kalas (1998), and suggested by the prototype Centaur, (2060) Chiron  (Kowal et al.\ 1979), whose dynamics are dominated by perihelic and aphelic interactions with giant planets.  We define the Centaurs as bodies whose orbits meet the
following conditions:

\begin{itemize}
\item The perihelion distance, $q$, and the semimajor axis, $a$, satisfy $a_J < q < a_N$ and  $a_J < a < a_N$, respectively, where $a_J$ = 5.2 AU is the semimajor axis of Jupiter and $a_N$ = 30.0 AU is the semimajor axis of Neptune.
 
\item The orbit is not in 1:1 mean-motion resonance with any planet.
\end{itemize}

This definition has the practical advantage of simplicity; it efficiently isolates a sub-set of the Solar system small-body population whose members are short-lived by virtue of their gravitational interactions with the giant planets.     At the time of writing there are, by the above definition, 92 known Centaurs.  

In this paper, we focus attention on Centaurs which display comet-like mass-loss.  Our purpose is to identify common properties of the activity and to seek evidence concerning its origin.  The active Centaurs have, so far, escaped systematic study as a group.  One likely reason is that the cometary classification afforded to such objects makes no distinction between them and other, long-studied comets in the Jupiter family, Halley family and Long-Period populations.   We seek answers to such questions as 1) How do the properties of active Centaurs compare with those of other comets and with Centaurs as a whole? 2) Is mass-loss implicated in the reportedly bimodal distribution of the optical colors of the Centaurs? 3) What drives the mass-loss from the active Centaurs? 

\section{Observations} 
The imaging observations reported here were taken at the University of Hawaii 2.2-m telescope (UH 2.2-m) and the Keck 10-m telescope, both located atop Mauna Kea, Hawaii.  At the UH 2.2-m we employed a Tektronix 2048$\times$2048 pixel charge-coupled device (CCD) camera having an image scale of 0.219 arcsec (pixel)$^{-1}$ when mounted at the f/10 Cassegrain focus.  At the Keck, we employed the Low Resolution Imaging Spectrograph (LRIS) which, in its modern form, employs independent red and blue channels separated by a dichroic beam-splitter (Oke et al.\ 1995).  A 5000~\AA~dichroic was used. The red-side detector is a front-side illuminated Tektronix 2048$\times$2048 pixel CCD camera, with image scale 0.210 arcsec (pixel)$^{-1}$ and two read-out amplifiers.  The blue-side used two Marconi (E2V) CCDs, each 2048$\times$4096 pixels in size with two readout amplifiers per chip, and with a scale of 0.135 arcsec (pixel)$^{-1}$.    On the blue side we used a B filter with effective wavelength $\lambda_e$ = 4370\AA~and full-width at half maximum (FWHM) of $\Delta \lambda$ = 900\AA.  On the red side, the filters employed were V 
($\lambda_e$ = 5473\AA, $\Delta \lambda$ = 950\AA), R ($\lambda_e$ = 6800\AA, $\Delta \lambda$ = 1270\AA) and I ($\lambda_e$ = 8330\AA, $\Delta \lambda$ = 3130\AA).

At both telescopes, we constructed flats by taking dithered images of the twilight sky and then 
median combining these images, after subtracting a bias frame and scaling them to a common
value, to eliminate stars.  For comparison, we also constructed ``data flats'' from the median combination of images
of astronomical targets taken through the night, relying on the lack of spatial correlation between images to eliminate field stars.  We found that the data-flats were superior to the twilight-sky flats in the B and I filters and we used them exclusively.  In V and R the two types of flat were comparable in quality. 
Photometric calibration of the flattened images was obtained from measurements of Landolt standard stars (Landolt 1992).  We used standard stars with optical colors similar to, or slightly redder than, the
Sun, in order to minimize color terms.  The faintest Landolt stars were used in order to minimize any possible shutter errors, and we experimented with different exposures to confirm that the shutter timing response was linear at the 1\% (0.01 mag.) level.  Lastly, we avoided standard stars with listed photometric uncertainties greater than 0.01 mag.

We observed a total of 23 Centaurs. A journal of observations is given in Table \ref{journal} and the main orbital parameters of these Centaurs are listed in Table \ref{orbits}.  Not all objects listed in Table \ref{journal} appear in subsequent Tables because of field-star contamination and other problems that caused us to reject some data from further consideration.  The estimated semimajor axes of two of the objects in our sample (namely, (148975) 2001 XA255 with $a$ = 30.183 AU and 2002 PN34 with $a$ = 30.781) have been revised following the acquisition of observations, such that they (slightly) exceed the semimajor axis of Neptune (30.070 AU).  We have elected to keep them in the sample, while noting that none of the conclusions drawn would be different if we instead chose to exclude them.

The next section gives brief introductions to the active Centaurs considered in this work.

\subsection{29P/Schwassmann-Wachmann 1}
\begin{figure}[]
\begin{center}
\includegraphics[width=0.8\textwidth]{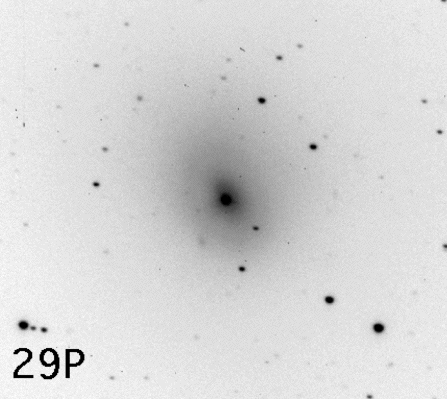}
\caption{29P/Schwassmann-Wachmann 1 imaged in the R-band filter on UT 2002 Sep 08.  This is a
300 sec integration with the University of Hawaii 2.2-m telescope.  The telescope was tracked non-sidereally
to follow the motion of the Centaur.  Field shown has North to the bottom, East to the right, and is
130 arcsec wide.  The centrally located Centaur is marked with a bar.  \label{29P}   } 
\end{center} 
\end{figure}

This object is labeled, and traditionally regarded, as a comet but it also meets the dynamical definition of a Centaur and so it is appropriate to consider it here.
Our optical observations (Table \ref{deltam}) show 29P in an active state, with a visible extent much larger than 100 arcsec (Figure \ref{29P}).
The colors are slightly redder than the Sun and essentially constant with respect to aperture size, except for a marginally significant trend towards bluer colors at
larger radii.    Earlier observers have reported that the broadband colors of 29P measured within a very large (88 arcsec diameter) aperture vary with the level of activity, becoming redder as the comet brightens (Kiselev and Chernova 1979).   

Since discovery in 1927, 29P has shown repeated photometric outbursts superimposed upon a platform
of constant activity (e.g.\ Roemer 1958, Jewitt 1990, Trigo-Rodriguez et al.\ 2008).  Spectroscopic measurements show that the activity is driven
by the sublimation of carbon monoxide (CO) ice (Senay and Jewitt 1994; Crovisier et al.\ 1995), with CO mass-loss rates of 1000 to 2000 kg s$^{-1}$.  Other species have been reported when the comet is
at its brightest, including CO$^{+}$ and CN (Cochran and Cochran 1991).  

Thermal measurements give an upper limit to the nucleus radius of 27$\pm$5 km and a lower limit to the geometric albedo 
of $p_R$ = 0.025$\pm$0.01 (Stansberry et al.\ 2004).  (These values are limits because of the substantial contamination of the
infrared data by coma, and the need to make a model-dependent correction for coma.)

\subsection{39P/Oterma}
\begin{figure}[]
\begin{center}
\includegraphics[width=0.8\textwidth]{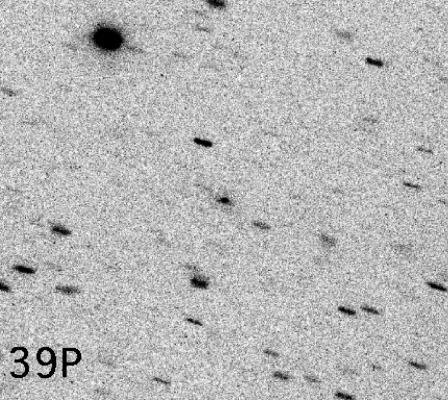}
\caption{39P/Oterma imaged in the R-band filter on UT 2002 Sep 08.  This is the sum of
two consecutive 900 sec integrations with the University of Hawaii 2.2-m telescope.  The telescope was tracked non-sidereally
to follow the motion of the Centaur.  Field shown has North to the bottom, East to the right, and is
130 arcsec wide.   \label{39P}   } 
\end{center} 
\end{figure}

The orbital parameters of this Centaur evolve substantially on timescales of decades owing to close interactions with Jupiter.  
For example, the perihelion was raised from 3.39 AU to 5.47 AU by a close approach (0.1 AU) to Jupiter in
1963, after which 39P was not re-observed until 1998.    In our data from the UH 2.2-m telescope taken 2002 Sep 07, 
39P was found to show a weak coma, barely visible in Figure \ref{39P}.  
The object is slightly redder than the Sun, with no
compelling evidence for a color gradient within the apertures employed for photometry (Table \ref{newcolors}).

\subsection{165P/Linear}
Coma was reported around this object (formerly known as 2000 B4) in observations taken with a 0.65-meter diameter telescope on UT 2000 Feb 10 (Kusnirak and Balam 2000) but no coma has been reported since this initial report.  Our own attempts to observe the object were made on UT 2002 Jan 09 using the UH 2.2-m telescope and on UT 2003 Jan 08, using the Keck 10-m telescope.  The object was not detected down to approximate limiting magnitudes $m_R$ = 23 in the former case and 24.5 in the latter.   We conclude that either the object was very faint or the ephemeris position was in error by more than the field radius (about 3.5 arcminutes).   Object 165P/Linear will not be further discussed in this paper.    

\subsection{166P/2001 T4}

\begin{figure}[]
\begin{center}
\includegraphics[width=0.8\textwidth]{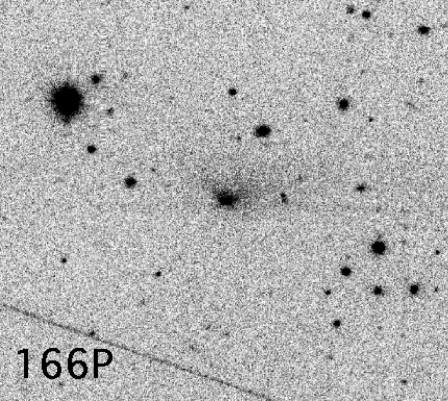}
\caption{166P/2001 T4 imaged in the R-band filter on UT 2002 Sep 07.  This is a
300 sec integration with the University of Hawaii 2.2-m telescope.  The telescope was tracked non-sidereally
to follow the motion of the Centaur.  The diagonal line is from a passing satellite.  Field shown has North to the bottom, East to the right, and is
130 arcsec wide.   \label{166P}   } 
\end{center} 
\end{figure}

Our imaging data show a clear coma at all epochs (Figure \ref{166P}) and the $BVRI$ color data show that 166P is optically very red.  For example, in
2001 Nov (Table \ref{newcolors}) we find B-V = 0.87, V-R = 0.69, R-I = 0.78 within a 5 pixel (1.1 arcsec) radius aperture.   Bauer et al.\ (2003) reported 166P to be one of the reddest objects in the Solar system, with V-R = 0.95$\pm$0.04 and R-I = 0.69$\pm$0.07 measured in 2001 Oct and 2002 Jan within a 1.5 arcsec radius aperture.   The reason for the difference
is not clear: we used the same telescope and filters and our measurements were obtained within a month of Bauer et al.'s first
observations.  These authors also reported a radial color gradient in 166P, with the image core being substantially redder than the
coma.  Our data confirm the existence of such a radial color gradient (Table \ref{newcolors}).  

\subsection{167P/2004 PY42}
\begin{figure}[]
\begin{center}
\includegraphics[width=0.8\textwidth]{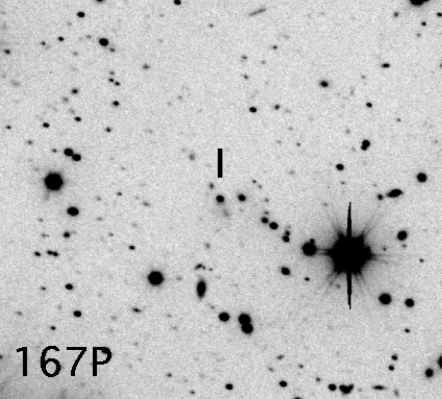}
\caption{167P/2004 PY42 imaged in the R-band filter on UT 2002 Sep 08.  This is the sum of two
200 sec integrations with the Keck 10-m telescope.  The telescope was tracked non-sidereally
to follow the motion of the Centaur.  Field shown has North to the bottom, East to the right, and is
130 arcsec wide.   \label{167P}   } 
\end{center} 
\end{figure}

This Centaur was observed on UT 2004 Oct 10 at the Keck 10-m telescope under photometric conditions.  A weak coma
was apparent in the raw data, and can be seen in Figure \ref{167P} (note that the diffuse object to the north east of 167P in
the figure is a background galaxy).  The coma is faint compared to the nucleus.  

\subsection{C/2001 M10 (Neat)}
\begin{figure}[]
\begin{center}
\includegraphics[width=0.8\textwidth]{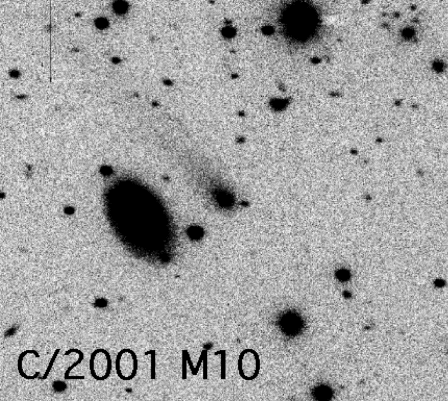}
\caption{C/2001 M10 imaged in the R-band filter on UT 2002 Sep 08.  This is a
300 sec integration with the University of Hawaii 2.2-m telescope.  The telescope was tracked non-sidereally
to follow the motion of the Centaur. Field shown has North to the bottom, East to the right, and is
130 arcsec wide. \label{M10}   } 
\end{center} 
\end{figure}

This Centaur was observed on UT 2002 Sep 08 at the UH 2.2-m telescope and on UT 2004 Oct 10 using the Keck 10-m telescope (Figure \ref{M10}).  As a result of instrumental problems, observations on the former date were obtained only in the $B$ and $R$ filters, while on the latter date full $BVRI$ colors were secured.  The object faded by
$\sim$3.5 magnitudes between 2002 and 2004, as the heliocentric distance increased from 6.056 AU to 9.145 AU.  Only 
$\sim$2 magnitudes of this fading can be attributed to geometric effects, the remaining $\sim$1.5 mag. reflects a real decrease in the coma
cross-section of C/2001 M10.  However, the $B-R$ color indices at the two epochs of observation, 1.22$\pm$0.08 and 1.12$\pm$0.09, are completely
statistically consistent within the measurement uncertainties.  Since the coma contribution to the signal in the UT 2004 Oct 10 data was reduced relative to that from the nucleus
by a factor $\sim$4, this suggests either that the nucleus contributes a negligible fraction of the total cross-section even at 9 AU or that the
nucleus and the coma have the same $B-R$ color.

\subsection{P/2004 A1 (Loneos)}
\begin{figure}[]
\begin{center}
\includegraphics[width=0.8\textwidth]{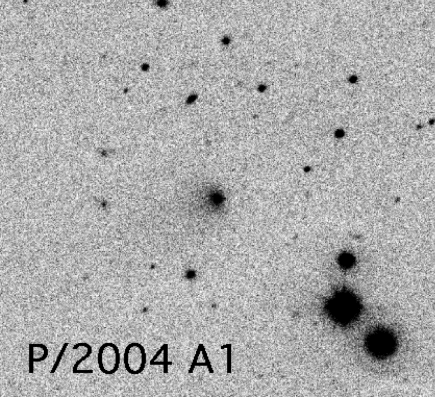}
\caption{P/2004 A1 (Loneos) imaged in the R-band filter on UT 2006 Jul 01.  This is a
300 sec integration with the University of Hawaii 2.2-m telescope.  The telescope was tracked non-sidereally
to follow the motion of the Centaur.  Field shown has North to the bottom, East to the right, and is
130 arcsec wide. \label{2004A1}   } 
\end{center} 
\end{figure}

This object passed within 0.03 AU of Saturn in 1992 causing a sudden decrease in the perihelion distance from $\sim$10 AU to
$\sim$5 AU (Hahn et al.\ 2006).  We observed it post-perihelion on UT 2006 July 01 at $R$ = 6.079 AU, when it displayed a prominent
coma (Figure \ref{2004A1}).   The absolute magnitude we derived, $m_R(1,1,0) >$ 10.92, is consistent with the faintest values reported by
Mazzotta et al. (2006).



\subsection{174P/Echeclus (60558)}
\begin{figure}[]
\begin{center}
\includegraphics[width=0.8\textwidth]{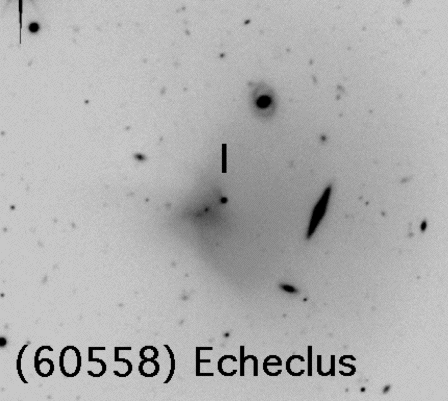}
\caption{174P/Echeclus (60558) imaged in the R-band filter on UT 2005 Feb 25.  This is a
300 sec integration with the Keck 10-m telescope.  The telescope was tracked non-sidereally
to follow the motion of the Centaur.  Field shown has North to the bottom, East to the right, and is
130 arcsec wide. The primary nucleus is marked with a bar.   \label{EC98}   } 
\end{center} 
\end{figure}

The object (formerly asteroid 2000 EC98) was discovered and measured as an inert Centaur (Rousselot et al.\ 2005) but promptly erupted into activity in late 2005 (Choi and Weissman 2006).  We imaged the object using Keck on 2006 February 25 (Figure \ref{EC98}).  The coma showed a complicated post-outburst structure, visible up to
65 arcsec from the nucleus.  We digitally removed 
background stars from the images and determined magnitudes within the standard apertures.  The magnitude of the nucleus corrected to unit heliocentric and geocentric distances and to zero degrees phase angle, $m_R(1,1,0) > 8.7$ (Table \ref{deltam}), is consistent with the pre-activity value, $H_R$ =  9.03$\pm$0.01, measured by Rousselot et al.\ Only $R$ filter color measurements were possible at Keck due to time constraints (Table \ref{newcolors}).  
However, spatially resolved colors were determined on UT 2006 Feb 24 by Bauer et al.\ (2008).  They found a coma slightly redder than sunlight with no substantial color gradients in the 5 arcsec to 30 arcsec projected radius range.

The nature of this object remains a mystery.  The motion of the active secondary component suggests that it is a fragment ejected from the parent body, but neither radiation forces nor non-gravitational forces due to asymmetric outgassing seem capable of producing the large relative velocity of separation (Weissman et al.\ 2006).    Thermal infrared measurements show that the coma particles are large and emitted continuously, not impulsively (Bauer et al.\ 2008).

\subsection{(5145) Pholus}
(5145) Pholus (formerly 1992 AD) is well known for its ultra-red optical reflection spectrum and presumed organic composition.  No coma has ever been reported around this object and no coma was detected in our Keck images taken UT 2004 Feb 17 (Table \ref{deltam}).

\subsection{(63252) 2001 BL41}
This object was observed as a point source on UT 2002 Feb 22 and 23 at the UH 2.2-m telescope with very similar apparent magnitudes on the two nights (Table \ref{deltam}).

\subsection{2003 QD112}
\begin{figure}[]
\begin{center}
\includegraphics[width=0.8\textwidth]{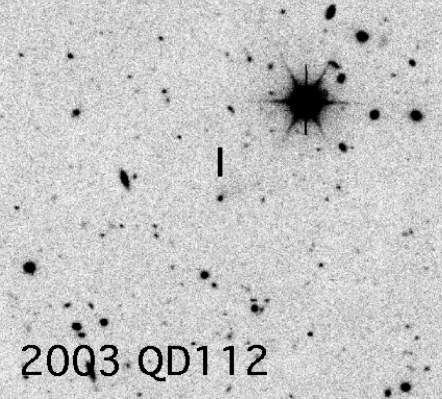}
\caption{2003 QD112 imaged in the R-band filter on UT 2004 Oct 10.  This is a
300 sec integration with the Keck 10-m telescope.  The telescope was tracked non-sidereally
to follow the motion of the Centaur.  Field shown has North to the bottom, East to the right, and is
130 arcsec wide. The centrally located Centaur is marked with a bar.  \label{QD112}   } 
\end{center} 
\end{figure}

We observed 2003 QD112 at the Keck telescope on UT 2004 Oct 10 when at 12.65 AU heliocentric distance and discovered a coma
about this object (Figure \ref{QD112}).  Only $R$-band data were secured (Table \ref{deltam}).  We reimaged
2003 QD112 at the Keck on UT 2007 Oct 12.  A coma was again evident in a blinked pair of 300-sec R-band integrations, but the comet
appeared in projection near to a much brighter field galaxy, making accurate photometry impossible on this date.  

\subsection{P/2005 S2 (Skiff)}
We attempted to image P/2005 S2 on UT 2007 Sep 19 using the UH 2.2-m telescope.  The object was not detected in 300-s $R$-band exposures with a limiting magnitude near $R$ = 23.5, leading us to conclude either that the brightness was less than expected or that the ephemeris was substantially in error.  Subsequently, we detected P/2005 S2 as a point-source object near magnitude 23.0 
(Table \ref{deltam}) using the Keck 10-m on UT 2007 Oct 12.

\subsection{P/2005 T3 (Read)}
This Centaur was not detected on either of two attempts (see Table \ref{journal}) with two different telescopes.  On UT 2007 Sep 17 the
object was projected against a background H$_2$ nebula which may have impeded its detection.  On UT 2007 Oct 12 a bright background star filled the field of view with scattered light.

\section{Observational Results}
\subsection{Coma Search}
We first visually searched each Centaur for coma by eye, comparing images taken at different times in order to distinguish faint near-nucleus structure from occasional background objects.  We also used multi-aperture photometry to measure, or set limits to, light from near-nucleus coma. Given a point-source nucleus embedded in an extended coma, the inner aperture of radius $\phi_1$ would sample the Centaur nucleus combined with an uncertain contribution from near-nucleus coma while the annulus between $\phi_1$ and an outer aperture of radius $\phi_2$ would provide a measure of pure coma.  Then, the apparent magnitude of the coma in the annulus between $\phi_1$ and $\phi_2$ is given by

\begin{equation}
m_{d} = -2.5 \log_{10}\left(10^{-0.4m_{2}} - 10^{-0.4m_{1}}\right)
\label{comamag}
\end{equation}

\noindent in which $m_1$ and $m_2$ are the magnitudes within apertures of radii $\phi_1$ and $\phi_2$.    In practice, a small correction is needed because the point-spread function (PSF) of the image is non-zero in the annulus between $\phi_1$ and $\phi_2$.  We measured the magnitude of this correction to the brightness in the coma annulus using the PSFs of field stars in the same images as the Centaurs.    

Based on trial and error, we selected aperture radii $\phi_1$ = 2.2 arcsec and $\phi_2$ = 3.3 arcsec as standard for our coma-search measurements.  Smaller values of $\phi_1$ proved overly sensitive to the steep surface brightness gradients found near the image core and to small differences in the PSF caused by relative motion between the Centaurs and the field stars used for reference.  Conversely, substantially larger apertures, $\phi_2$, were more frequently contaminated by faint background sources and more subject to photometric errors from the uncertainty in the brightness of the night sky.    With this choice of $\phi_1$ and $\phi_2$, we found that the wings of the PSF between $\phi_1 \le \phi \le \phi_2$  typically contained only $\sim$1\% to 2\% of the light with $\phi < \phi_1$.  

The resulting coma annulus magnitudes, $m_d$, are listed in column 5 of Table \ref{deltam}.  Where no significant coma is detected, we have listed a 3$\sigma$ limit to $m_d$ computed from the scatter in multi-annulus measurements of $\sim$5 to $\sim$10 field stars.  Coma is detected in nine of the 23 Centaurs observed.  In eight of these nine cases, coma that was detected through multi-aperture photometry and Equation (\ref{comamag}) was also visible to the eye in highly stretched images blinked rapidly on a computer.   Only in Centaur 2006 SX368 did we find evidence for coma from the photometry that could not be corroborated visually.  

\subsection{Absolute Magnitudes and Sizes}
The magnitudes of the Centaurs are corrected to $R_{au}$ = $\Delta_{au}$ = 1 and phase angle $\alpha$ = 0 using

\begin{equation}
m_R(1,1,0) = m_R - 5 \log_{10}(R_{au} \Delta_{au}) - \Phi(\alpha)
\label{absolute}
\end{equation}

\noindent where $R_{au}$ and $\Delta_{au}$ are the heliocentric and geocentric distances in AU and $\Phi(\alpha)$ is the so-called phase function correction at phase angle $\alpha$.  The phase angles, $\alpha$, at which the Centaurs were observed span the range from 0 $\le \alpha \le$ 10$^{\circ}$ (Table \ref{journal}).  To first order, the phase function over this modest range of angles can be represented as having a linear (in magnitude space) dependence on $\alpha$, given by  $\Phi(\alpha)$ = 10$^{(-0.4\beta \alpha)}$, where $\beta$ [mag. deg$^{-1}$] is a constant.  However, at small phase angles, the phase functions of many objects display a narrow-angle peak (the ``opposition-surge'') while at larger phase angles the phase function often appears measurably non-linear.  The phase functions of the active Centaurs are further uncertain, because the angle dependence of the scattering from the dust particles in the coma is not necessarily like that of the solid surface of the nucleus from which those particles were ejected.  Here, we used the ``HG'' formalism of Bowell et al.\ (1989) with scattering parameter $G$ = 0.15.  This choice gives phase darkening similar to the constant $\beta$ = 0.1 mag. deg$^{-1}$ case (e.g.\ the Bowell formalism gives phase darkening at $\alpha$ = 10$^{\circ}$ to 35\% of the opposition value while $\beta$ = 0.1 mag. deg$^{-1}$ gives a 40\% darkening).   

The absolute magnitudes, $m_R(1,1,0)$, are listed in Table \ref{deltam}.  The Table also lists effective object radii, $r_e$ [km], computed under the assumption of geometric albedo $p_R$ = 0.1.  The statistical uncertainty on $m_R(1,1,0)$ is the same as on $m_R$ but the actual uncertainty is larger, owing to uncertainty introduced by the phase angle correction.  The uncertainty on $r_e$ is potentially large and unknown, pending future determinations of the albedos and phase functions of the observed objects.  We include $m_R(1,1,0)$ and $r_e$ here mainly to provide an approximate indication of the relative sizes of the observed objects.  Note that entries for active Centaurs in Table \ref{deltam} are presented as lower and upper limits on $m_R(1,1,0)$ and $r_e$, respectively, with no attempt to remove coma from the central aperture using profile-fitting or other techniques.   The smallest observed Centaurs, P/2005 S2 ($r_e$ = 2 km), 2000 GM137 ($r_e$ = 3 km) and 39P ($r_e <$ 3 km) have sizes consistent with the nuclei of better-observed Jupiter family comets.  However, the median radius of the observed (inactive) Centaurs is $r_e$ = 15 km (Table \ref{deltam}), about an order of magnitude larger than the characteristic radii of well-measured Jupiter family comet nuclei (Lamy et al.\ 2004) and an order of magnitude smaller than the characteristic radii of Kuiper belt objects for which reliable physical data exist.  These differences simply reflect the relative mean distances of the different types of object and the limitations of magnitude-limited photometry.

\subsection{Colors}
Like the Kuiper belt objects, the Centaurs are remarkable for their wide
range of optical colors, ranging from nearly neutral to very red.  In fact, two of the most famous Centaurs
span the full range of colors present in the outer Solar system: (2060) Chiron is nearly 
neutral (Meech and Belton 1990, Luu and Jewitt 1990) and shows weak bands of water ice (Foster et al.\ 1999, Luu et al.\ 2000)
while (5145) Pholus is very red (Mueller et al.\ 1992), perhaps because of
the presence of irradiated complex organic compounds (Luu et al. 1994, Cruikshank et al.\ 1998).
The fact that Chiron is an active Centaur while Pholus is not has lead several investigators to
speculate that outgassing activity might be responsible for the neutral colors of some Centaurs.  For example, outgassing
could blanket the surface of a body with neutral-colored fine particles excavated from beneath the few
meter-thick surface layer in which radiation damage is strong (Delsanti et al.\ 2004).
There is some evidence that the Centaurs as a whole occupy a bimodal distribution of optical 
colors (Peixinho et al.\ 2003).   If so, the bimodality could tell us which Centaurs have been recently active.  However, other interpretations are possible.  Perhaps activity does not affect the surface colors.  The color differences could reflect different intrinsic compositions, possibly related to different Centaur source regions.

\begin{figure}[]
\begin{center}
\includegraphics[width=0.75\textwidth]{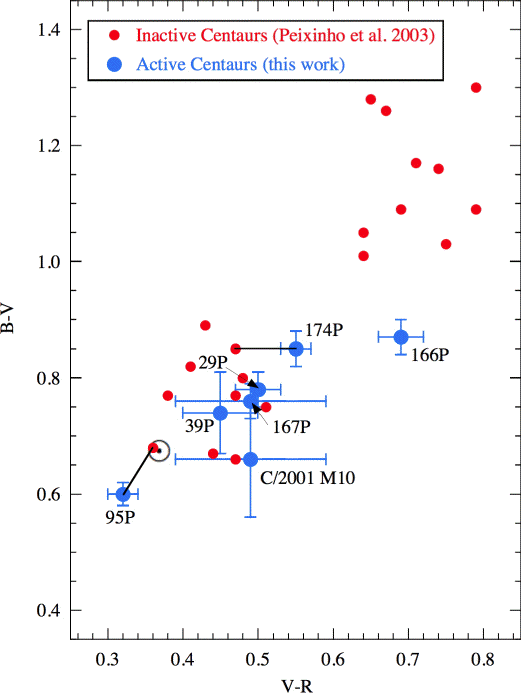}
\caption{The B-V color index is plotted against the V-R index, using data from Table \ref{finalcolors} for the active Centaurs (blue circles) and from Peixinho et al.\ 2003 for the
inactive Centaurs (red circles).  Objects present in both samples are
connected by straight lines.   The color of the Sun is marked.   \label{BVvsVR}   } 
\end{center} 
\end{figure}

Figure \ref{BVvsVR} shows the distribution of the Centaurs in the B-V vs. V-R color plane.  The data on 
inactive Centaurs (red circles) are taken from the compilation of Peixinho et al.\ (2003), while data on active Centaurs (blue circles with error bars) are from 
Table \ref{finalcolors}.  Measurements of two Centaurs (Chiron and 60558) appearing in both the Peixinho et al.\ 
compilation and the present work are joined by lines.  For clarity, only the uncertainties on the photometry of the active Centaurs are shown.  The Figure shows the ``red'' and ``blue'' color clumps 
occupied by the Centaurs in the Peixinho sample. Six of the active Centaurs for which we possess quality B-V and V-R colors fall within the ``blue'' clump, while the seventh, 166P, is red in V-R but not in B-V, and falls in neither the blue clump nor the red. 

\begin{figure}[]
\begin{center}
\includegraphics[width=0.8\textwidth]{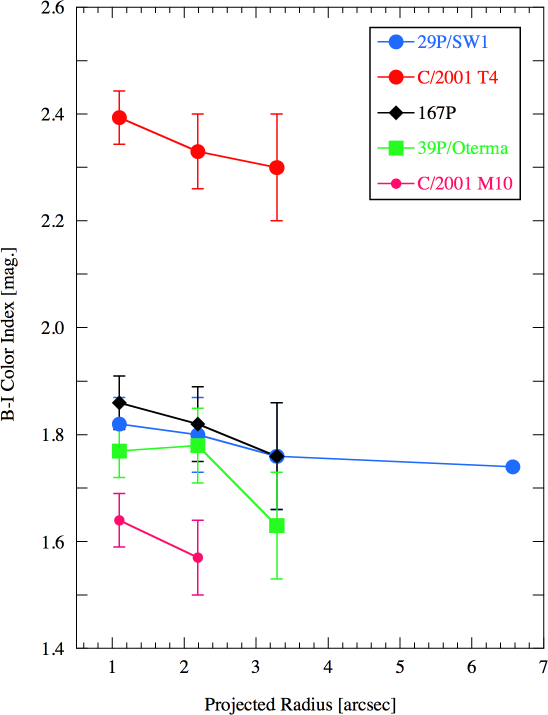}
\caption{Average $B-I$ color index vs. projected radial distance from the photocenter for five active Centaurs.
General trend toward bluer colors at larger radii is apparent.  \label{BIvsradius}   } 
\end{center} 
\end{figure}

An observation relevant to the more neutral colors of the active Centaurs is provided by Figure \ref{BIvsradius}, which shows the $B-I$ color index measured as a function of the projected 
angular distance from the photocenter of the five active Centaurs. Only these five objects have adequate spatial data and
signal-to-noise ratio to justify an examination of the radial color gradient.  The wide range in the $B-I$ colors of the active Centaurs is apparent from
this Figure, from the very red 166P/2001 T4 to nearly neutral C/2001 M10 (the Solar color index is 
$(B-I)_{\odot}$ = 1.38).  The main trend apparent 
in Figure \ref{BIvsradius} is the tendency for the $B-I$ color index to decrease with angular radius, meaning that the comae 
of the active Centaurs are typically bluer than their centers.  We used the profiles of field stars to confirm that the color gradients in Figure (\ref{BIvsradius}) could not be artifacts
produced by differences in the image quality.  A radial color gradient in the observed sense has been observed
for some active Jupiter family comets.  For example, P/Tempel 2 had quite different nucleus and coma colors of V-R = 0.55$\pm$0.03 and 0.36$\pm$0.03, respectively (A'Hearn et al.\ 1989, Jewitt and Luu 1990).  Relatively blue coma is plausibly interpreted as an effect of non-geometric
scattering in optically small particles, so perhaps Figure \ref{BIvsradius} simply shows that the particles ejected from the active
Centaurs are, on average, small enough for geometric effects to imbue them with a color different from the underlying nucleus material.  More complex spatial gradients have been previously reported in 29P/SW1, where they may be associated with particle size sorting by solar radiation pressure (Cochran et al.\ 1982).  Unfortunately, no detailed study of color gradients in cometary comae has been published, and the relation between the coma color and the color of the underlying nucleus cannot be specified with confidence.  

In summary, the active Centaurs tend to lack the ultrared matter observed in the Kuiper belt and some inactive Centaurs.  We find no evidence that the colors of the active Centaurs are bimodally distributed (Figure \ref{BVvsVR}) but the number of well-measured objects is small. The colors of the active Centaurs might reflect a true lack of ultrared matter on their nuclei, or simply an artifact of color dilution with neutral-blue coma particles in the near-nucleus environment (Figure \ref{BIvsradius}).  In view of the complications imposed by coma contamination of the near-nucleus photometry, we are reluctant to draw any strong conclusions based on the correlation between Centaur color and activity.

\subsection{Orbital Distribution of the Active Centaurs}

\begin{figure}[]
\begin{center}
\includegraphics[width=0.8\textwidth]{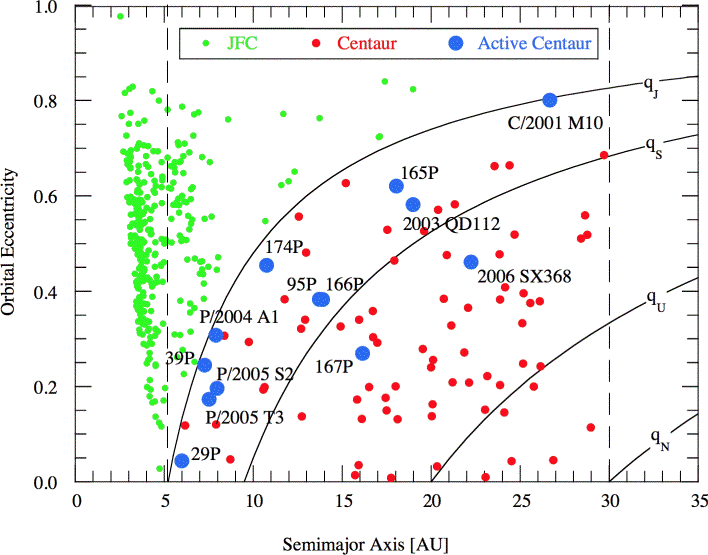}
\caption{Orbital semimajor axis vs.\ eccentricity for active (blue) and inactive (red) Centaurs, and for Jupiter
Family Comets (green).  Labeled curves
show the loci of orbits having perihelia equal to the semimajor axes of the four giant planets. Vertical
dashed lines show the semimajor axes of Jupiter and Neptune, bounding the region of the Centaurs.  \label{ae}   } 
\end{center} 
\end{figure}

Figure \ref{ae} shows the distribution of the Centaurs in the semimajor axis, $a$, vs. eccentricity, $e$, plane.  Active Centaurs, including those observed in the present sample plus others reported in the literature as being active,  are shown as blue circles, while all other Centaurs are 
shown in red.  Also plotted (green circles) are the Jupiter family comets (JFCs) defined as having a Tisserand parameter with 
respect to Jupiter 2 $\le T_J \le$ 3.  Lines are plotted in Figure \ref{ae} to mark the loci of orbits having perihelion distances equal to the semimajor axes 
of the giant planets.  Objects above the line marked ``$q_J$'', for instance, have perihelion inside Jupiter's orbit.  

\begin{figure}[]
\begin{center}
\includegraphics[width=1.0\textwidth]{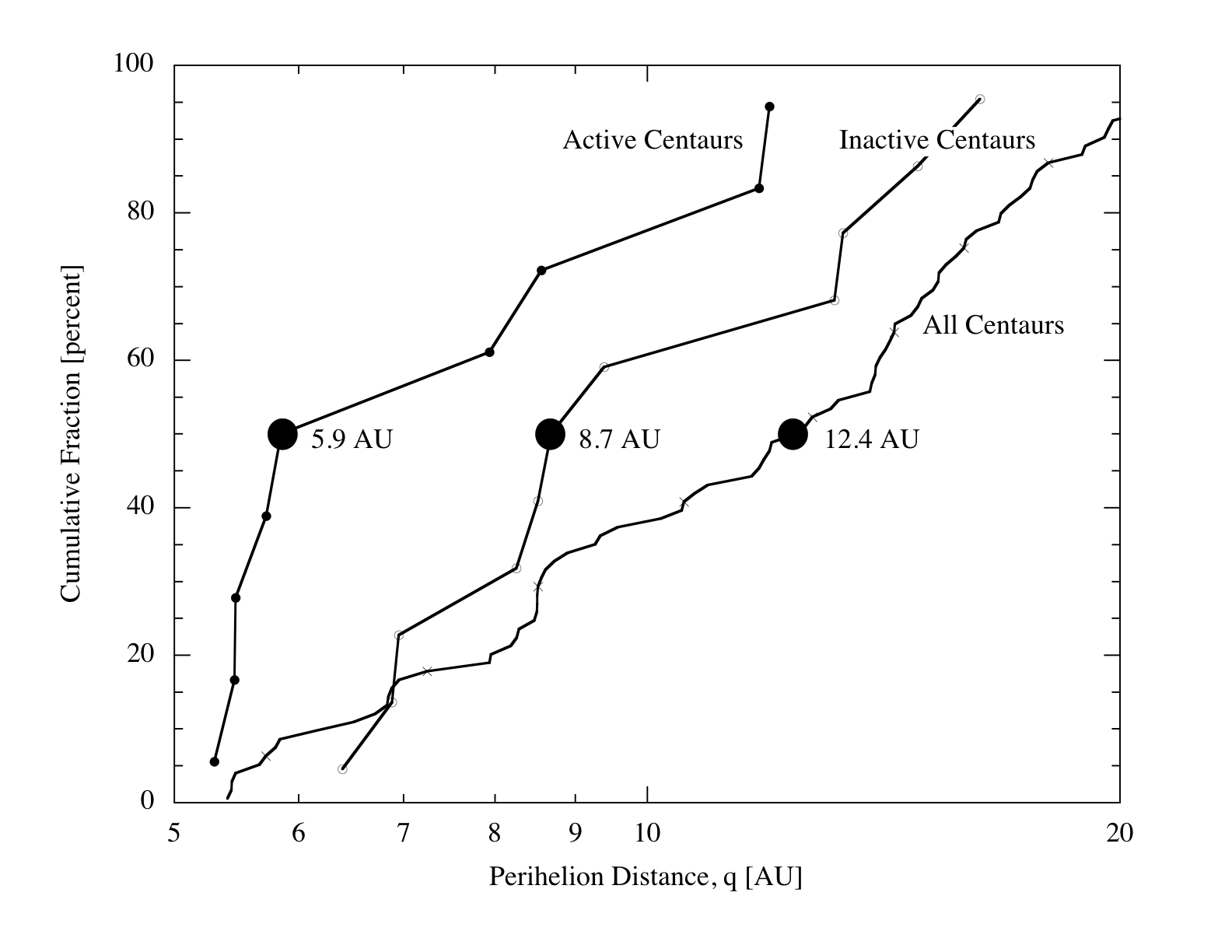}
\caption{The cumulative distributions of the perihelion distances of, from left to right, the active Centaurs in our sample, the inactive Centaurs in our sample and all known Centaurs.  The median perihelion distances for each group are marked.
\label{qfrac090105}   } 
\end{center} 
\end{figure}

Figure \ref{ae} shows that the distribution of the active Centaurs in the $a$ vs. $e$ plane is different from that of the Centaurs as a whole.  The active Centaurs tend to possess smaller than average perihelia, with most located in a band between the orbits of Jupiter (curve labelled $q_J$ in Figure \ref{ae}) and Saturn ($q_S$).   The median (mean) perihelion distance of the active Centaurs in our sample is 5.9 AU (7.6$\pm$0.9 AU, $N$ = 9 data points) whereas the corresponding quantities for the inactive Centaurs in our sample are 8.7 AU (10.2$\pm$1.1 AU, $N$ = 11).  The Centaurs as a whole have median (mean) perihelion distances 12.4 AU (12.5$\pm$0.5 AU, $N$ = 89). The difference is emphasized in Figure \ref{qfrac090105}, which shows the cumulative distributions of the perihelion distance for the active and inactive Centaurs and for the total Centaur population.  According to the Kolmogorov-Smirnov test, the statistical likelihood that the active and inactive Centaur perihelion distance distributions could be drawn by chance from the same parent population is 0.03, corresponding to about a 2.2$\sigma$ significance.  The corresponding numbers for the active Centaurs and the total Centaur population are 0.01 and $\sim$2.7$\sigma$.

Could these differences be artifacts of an observational bias?  For example, is it possible that Centaurs of all perihelion distances are active but that the mass-loss can be detected only in those with the smallest $q$?  To investigate this possibility, we must understand the relation between the detection of a coma and the rate of mass-loss from the nucleus responsible for that coma.

\subsection{Mass-Loss Rates}

We seek to relate the optical photometry to the mass production rates of the Centaurs.  To do this, we employ a three-step procedure. First, the photometry is used to estimate the scattering cross-section of the dust in the coma of each object.   Second, the cross-section is related to the dust mass through the adoption of a size distribution of dust particles that is based on independent measurements from comets near 1 AU.  Third, the mass-loss rate is inferred by dividing the dust mass by the time of residence of the dust grains as they travel across the projected photometry aperture.  There are many unknowns in each step of this procedure and the derived mass-loss rates are, at best, crude estimates of the true values.  Nevertheless, in the absence of more information, this is the best we can do.

Magnitude $m_d$ (Equation \ref{comamag}) is related to the geometric albedo, $p_R$, and total cross-section of the coma dust particles within the projected annulus, $C_d$,  by the inverse square law,  conventionally expressed (Russell 1916) as

\begin{equation}
p_R \Phi(\alpha)C_d = 2.25\times10^{22}\pi R_{au}^2\Delta_{au}^2 10^{-0.4(m_d - m_{\odot})}.
\label{inversesq}
\end{equation}

\noindent Here, $R_{au}$ and $\Delta_{au}$ are the heliocentric and geocentric distances expressed in AU, $\alpha$ is the phase angle and $m_{\odot}$ is the apparent
magnitude of the Sun viewed from 1 AU.  The quantity $\Phi(\alpha)$ represents the phase darkening.  
We further take $p_R$ = 0.1, consistent with measurements of active comets in backscattering geometry (Kolokolova et al.\ 2004).    Equations \ref{comamag} and \ref{inversesq} together give the dust cross-section for each Centaur, which we list in Table \ref{massloss}.   The dust cross-section and the dust mass are related through the size distribution and other parameters of the dust (see Appendix).  We employ Equation (\ref{mass}) to relate $C_d$, the cross-section derived from the photometry to $M_d$, the dust mass. Values of $M_d$ are  also listed in Table \ref{massloss}.

The time of residence of the dust grains, at heliocentric distance $R_{au}$, in the annulus defined by $\phi_1$ and $\phi_2$ is just 

\begin{equation}
\tau(R_{au}) = 1.5\times10^{11} \Delta_{au} (\phi_2 - \phi_1) /v(R_{au}), 
\label{tau}
\end{equation}

\noindent where  $\Delta \phi$ = $\phi_2 - \phi_1$, expressed in radians, and $v(R_{au})$ is the radial outflow 
speed of the dust grains from the nucleus.  The heliocentric variation of the dust velocity is uncertain but is known to decrease with increasing heliocentric distance, albeit slowly.  For example, measurements of expansion speeds in C/Hale-Bopp over the distance range 4 $\le R_{au} \le$ 14  (Biver et al.\ 2002) are compatible with

\begin{equation}
v(R_{au}) = v_0 \left[\frac{R_0}{R_{au}}\right]^{1/4}
\label{speed}
\end{equation}

\noindent where $v_0$ = 550 m s$^{-1}$ is the velocity at reference distance $R_0$ = 5 AU.  
Equation (\ref{speed}) likely gives an upper limit to the true speed, since the Hale-Bopp measurements refer to gas. Entrained dust will, in general, not be perfectly coupled to the gas and therefore should travel more slowly.
 
From Equations (\ref{tau}) and (\ref{speed}), the time taken for dust grains traveling radially to cross the annulus is written

\begin{equation}
\tau(R_{au}) = 1.5\times 10^{11} \frac{ \Delta_{au} (\phi_2 - \phi_1) }{v_0} \left[\frac{R_{au}}{R_0}\right]^{1/4}
\label{crossingtime}
\end{equation}

\noindent This gives a simplistic estimate both because the grains do not radially cross the projected photometry annulus and because the grains presumably occupy a broad distribution of sizes, with the larger grains less efficiently coupled to the expanding gases of the coma and so traveling more slowly than small grains.  Still, as a first approximation, Equation (\ref{crossingtime}) provides a useful estimate.  Column 7 of Table \ref{massloss} shows that, for our data, values of $\tau$ typically fall in the range (1 to 4)$\times$10$^4$ s, or about an hour.  Model mass-loss rates computed from Equations (\ref{mass}) and (\ref{crossingtime}) are listed in Table \ref{massloss}.

\begin{figure}[]
\begin{center}
\includegraphics[width=0.8\textwidth]{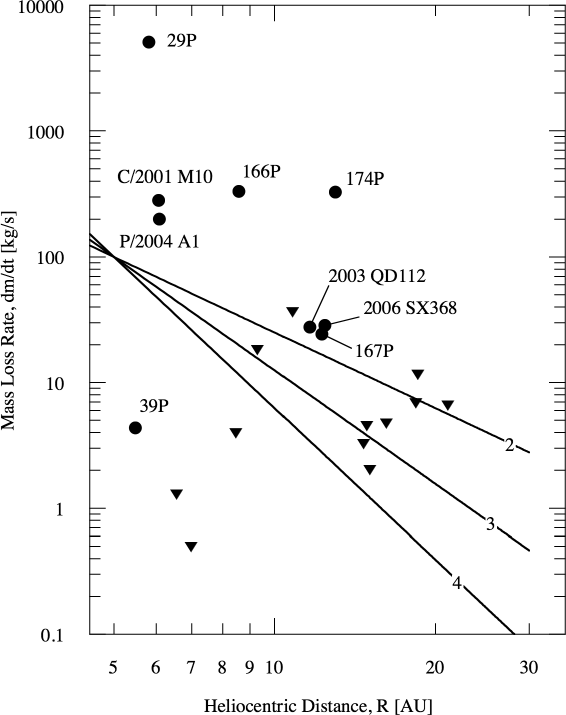}
\caption{Derived dust mass-loss rates vs. heliocentric distance for the Centaurs in this sample.  Circles represent measurements of the active Centaurs while inverted triangles represent upper limits to dust on Centaurs appearing inactive.  The solid lines show power-laws of slope $n$ = $2$, $3$ and $4$, as marked, for reference (see Equation (\ref{powerlaw})).  \label{dmbdtvsR}   } 
\end{center} 
\end{figure}

The derived mass-loss rates are plotted against heliocentric distance in Figure \ref{dmbdtvsR}, with filled circles to represent the active Centaurs and inverted triangles marking 3$\sigma$ limits to the mass-loss for those Centaurs in which no coma was detected.  In passing, we note that the mass-loss rates for 29P estimated from our dust measurements ($\sim$5000 kg s$^{-1}$) are of the same order as rates estimated independently, although at different times, from measurements of gas in this object ($\sim$1000 to 2000 kg s$^{-1}$: Senay and Jewitt 1994, Crovisier et al.\ 1995).  Mass loss rates from P/2004 A1 have also been independently estimated from imaging data by Mazzotta et al.\ (2006).  Their estimates ($\sim$100 kg s$^{-1}$) are within a factor of two of those in Table \ref{massloss} and Figure \ref{dmbdtvsR} ($\sim$200 kg s$^{-1}$).  We are aware of no other mass-loss rate estimates for the active Centaurs, but conclude from these limited comparisons that our simple photometric method produces values in reasonable accord with others.

The solid lines in Figure \ref{dmbdtvsR} show mass-loss rates varying as

\begin{equation}
\frac{dm}{dt} = 100 \left[\frac{5}{R_{au}}\right]^n
\label{powerlaw}
\end{equation}

\noindent where index $n$ = 2, 3 and 4 and the normalization to 100 kg s$^{-1}$ at $R_{au}$ = 5 is arbitrary. 
The Figure shows that those Centaurs observed to be active at $R_{au} <$ 13 would also be measurably active at larger distances, if their sublimation followed an $n$ = 2 dependence on $R_{au}$.  For example 166P, when displaced from 8 AU to 20 AU, would lose mass at about 60 kg s$^{-1}$, according to the Figure, which is an order of magnitude larger than the detection limits established for other Centaurs (e.g.\ 52975, see Table \ref{massloss}) at this distance.  [The one exception is the very weakly active 39P/Oterma: if displaced from 5.5 AU to larger distances an equilibrium coma around this object would quickly become unobservable].  Therefore, we conclude that the mass-loss rates in active Centaurs drop with heliocentric distance faster than $R_{au}^{-2}$. This is significant because $n$ = 2 is the index expected from the equilibrium sublimation of an exposed supervolatile ice (because essentially all the incident Solar energy is used for sublimation, leaving none for thermal re-radiation).  Solid carbon monoxide (CO), for instance, would sublimate in proportion to $R_{au}^{-2}$ if exposed to sunlight.   This suggests that the absence of coma around Centaurs with $R_{au} >$ 13 is not an artifact of observational selection acting upon a set of Centaurs losing mass in proportion to $R_{au}^{-2}$.    A radial variation with $n \ge$ 3 can fit the data in Figure \ref{dmbdtvsR} but this implies that the mass-loss is not driven simply by the sublimation of an exposed supervolatile.  It is interesting to note that observations of comet C/Hale-Bopp as it moved out through the giant planet region of the Solar system lend independent support to this conclusion.  The mass-loss rate from the comet closely followed $R_{au}^{-2}$ (Jewitt et al.\ 2008) and, indeed, coma has been consistently observed in optical images out as far as 26 AU (Szab{\'o}  et al.\ 2008).

On this basis we conclude that the activity in the Centaurs is not driven by the equilibrium sublimation of CO or another supervolatile.  Either the mass-loss is driven by a less volatile material (which would have $n >$ 2 in the observed range of heliocentric distances) or the restriction of the activity to smaller heliocentric distances is the result of a trigger which prevents mass-loss when far from the Sun.

\section{Discussion}
\subsection{Origin of the Activity}

The prime observational clues to the origin of Centaur activity are:

\begin{itemize}

\item \textbf{Perihelion distribution.} 
The median perihelion distance of the active Centaurs is 5.9 AU, whereas the median perihelion
distance for the Centaurs as a whole is 12.4 AU (Figure \ref{qfrac090105}).  As discussed above (see Figure \ref{dmbdtvsR}), it is unlikely that this difference is the result of observational selection acting on a set of active Centaurs in the strong sublimation limit.  Instead, the data suggest that the activity is triggered or driven by a temperature-related process.

\item  \textbf{Detection of CO.} Centaur 29P is a strong source of carbon monoxide, with a sustained outgassing rate 1000 kg s$^{-1}$ to 2000 kg s$^{-1}$ (Senay and Jewitt 1994; Crovisier et al.\ 1995, Gunnarsson et al.\ 2008).  CO has also been reported (but not confirmed) with a similar
outgassing rate in Centaur
(2060) Chiron (Womack and Stern 1999).  While other Centaurs show no spectral evidence
for CO sublimation or, indeed, for any other volatile (Bockelee-Morvan et al.\ 2001, Jewitt et al.\ 2008), these detections focus attention on the likely role of this supervolatile in driving mass-loss from the active Centaurs.

\item  \textbf{Line Profiles of CO.}  The rotational line profiles of CO in 29P are asymmetric, with a narrow, blue-shifted component (Senay and Jewitt 1994, Crovisier et al.\ 1995, Gunnarsson et al.\ 2008).  This shows that the release of the CO is diurnally modulated and therefore that the CO source is in good thermal contact with (i.e.\ close to) the physical surface of 29P.  The source cannot be deep.

\end{itemize}

We now expand upon these points.  

\begin{figure}[]
\begin{center}
\includegraphics[width=0.8\textwidth]{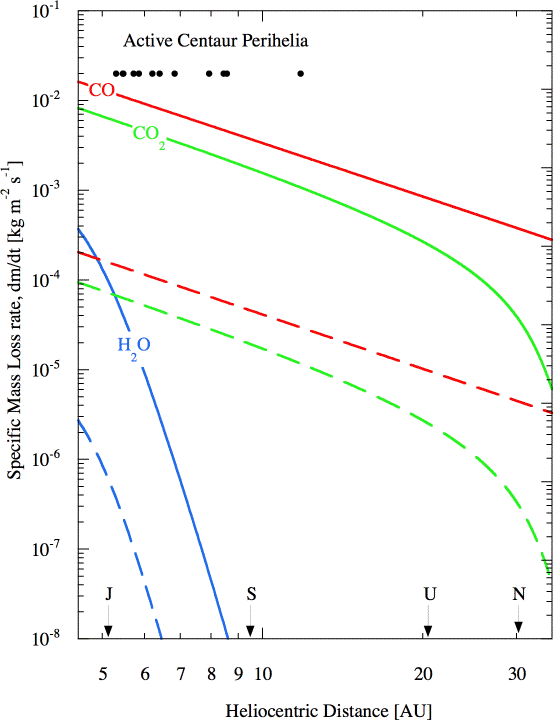}
\caption{(left axis) Specific mass-loss rate [kg m$^{-2}$ s$^{-1}$]  as a function of the heliocentric distance [AU] computed from Equation \ref{sublimation}.  Blue curves apply to crystalline water ice, green curves to CO$_2$ ice and red curves to CO ice.  Solid and dashed lines for each ice refer to the high and low temperature limiting cases, as described in Section \ref{sectionamorphous}. The radii of the orbits of the giant planets are marked for reference at the bottom of the plot.  The perihelion distances of the active Centaurs are marked at the top with black circles.  \label{dmbdt_plot}   } 
\end{center} 
\end{figure}

Figure \ref{dmbdt_plot} shows solutions to the energy balance equation:

\begin{equation}
\frac{L_{\odot} (1-A)}{4 \pi R^2} = \chi \left[\epsilon \sigma T^4 + L(T) \frac{dm}{dt} \right]
\label{sublimation}
\end{equation}

\noindent  in which $L_{\odot}$ = 4$\times$10$^{26}$ W is the luminosity of the Sun, $R$ is the heliocentric distance in meters,
$A$ is the Bond albedo, $\epsilon$ is the emissivity and $\sigma$ = 5.67$\times$10$^{-8}$ W m$^{-2}$ K$^{-4}$ is the Stefan-Boltzmann
Constant.  The second term on the right hand side accounts for sublimation at the rate $dm$/$dt$ [kg m$^{-2}$ s$^{-1}$], and
$L(T)$ [J kg$^{-1}$] is the (temperature dependent) latent heat of sublimation.  Conduction is neglected from Equation (\ref{sublimation}) based on the extremely small thermal inertias of two measured Centaurs (Fernandez et al.\ 2002) and of the nucleus of P/Tempel 1 (Groussin et al.\ 2007).  Quantity $\chi$ is a dimensionless parameter intended to represent the way in which the energy from the Sun is distributed over the surface of the body and is affected by both the thermal parameters of the nucleus and its rotational state.  Physically, $\chi$ may be interpreted as the ratio of the area from which absorbed sunlight is radiatively lost into space to the area over which energy is absorbed.  We considered two limiting cases to bracket the likely range of $\chi$ at any given heliocentric distance.  A flat surface exposed normally to the Sun absorbs and radiates from the same area and so has $\chi$ = 1.  This limit would apply, for example, to the subsolar point of a slowly rotating body or to any nucleus whose projected pole direction passed through the Sun.  At the other extreme, an isothermal sphere of radius $r$ absorbs over $\pi r^2$ but radiates from 4$\pi r^2$, and so has $\chi$ = 4.  
We took $L(T)$ from Brown and Ziegler (1979) and assumed $A$ = 0.1, $\epsilon$ = 0.9 (the particular values assumed do not have a critical effect on the computed temperatures or mass-loss rates provided $A \ll$ 1 and $\epsilon \gg$ 0).

The solutions to Equation (\ref{sublimation}) are shown in Figure \ref{dmbdt_plot} for crystalline water ice (blue lines), CO$_2$ ice (green lines) and CO ice (red lines).   For each ice we plot two curves to show the low and high $\chi$ values.  Furthermore, at each heliocentric distance we calculated the critical grain size, $a_c$, above which the gas drag force (proportional to $dm$/$dt$) is insufficient to launch a grain against the gravitational attraction to a nucleus of density $\rho$ = 10$^3$ kg m$^{-3}$ and radius $r_n$ = 25 km.  Figure \ref{dmbdt_plot} shows solutions of Equation \ref{sublimation} for which $a_c >$ 1 $\mu$m.  Grains much smaller than 1 $\mu$m are inefficient optical scatterers with characteristically blue colors that contradict the measured colors of the Centaurs.  Figure \ref{dmbdt_plot} shows that CO and CO$_2$ sublimation is able to eject 1 $\mu$m grains over the full range of heliocentric distances swept by the Centaurs while H$_2$O sublimation can only do so for $R_{au} \le$ 6.5 and 8.5, in the low and high temperature limits, respectively.

Figure \ref{dmbdt_plot} shows that none of the ices provide a convincing match to the Centaur activity.  The Centaurs are too cold, even at their subsolar points, for water sublimation to be effective except close to Jupiter's orbit.   For example, 2003 QD112, 166P, 167P and 2006 SX368 are active at 11.6, 8.6, 12.2 and 12.4 AU, all far beyond the water sublimation zone.  Other Centaurs such as C/2001 M10, P/2004 A1, 29P are active at distances (6.06-9.2 AU, 6.1 and 5.8 AU, respectively) where water ice can weakly sublimate, but a comparison of the specific mass-loss rates with the actual mass-loss rates (Table \ref{massloss}) shows that implausibly large areas of exposed water ice are required.  More importantly, spectroscopic observations clearly show that the mass-loss is controlled by large fluxes of CO, at least for 29P (Senay and Jewitt 1994, Crovisier et al.\ 1995).

Conversely, CO and CO$_2$ ices are so volatile that they should 
sublimate strongly throughout the planetary region.  If free CO ice were present on the Kuiper belt objects and Centaurs we should observe comae around these bodies at all heliocentric distances out to, and beyond, the orbit of Neptune.
Since this is clearly not the case, we conclude that free CO is not responsible for the observed activity.   Non-detections of CO rotational lines have been used to place empirical limits to the possible fractional areas of exposed CO on Centaurs and KBOs near 0.1 - 1\%, implying a strong depletion (Jewitt et al.\ 2008).  Physically, this conclusion is easy to understand.  Carbon monoxide ice sublimates so rapidly that any exposed CO 
could survive only briefly.  For example, even in the cold limit, CO sublimates at 30 AU at a rate near 
$dm$/$dt$ = 10$^{-5}$ kg m$^{-2}$ s$^{-1}$ (Figure \ref{dmbdt_plot}), corresponding to recession of the ice surface (for density $\rho$ = 1000 kg m$^{-3}$) at 10$^{-8}$ m s$^{-1}$.  In one orbital period of $\sim$150 yrs, 50 meters of CO would sublimate away.  In 1000 orbits (only $\sim$150kyr) the sublimation distance would rival the radius of the Centaur. Clearly, free CO ice cannot persist near the surface of any Centaur or Kuiper belt object and might only be found, if anywhere, at the bottom of deep, self-shadowing pits created by sublimation.  
The only exception to this statement applies to the largest Kuiper belt objects, such as Pluto, on which sublimated CO can be recycled through a thin, gravitationally bound atmosphere.    The much smaller objects that dominate the Centaur population must all be free of exposed supervolatile ices on their surfaces.

Could the CO be buried beneath a refractory layer and heated only slightly by the Sun?  In this way, mass-loss at large $R_{au}$ might be strongly suppressed relative to the $n$ = 2 heliocentric dependence expected of supervolatile ice.  The main argument against this possibility is provided by the profiles of the CO rotational lines recorded in submillimeter spectra.  The lines show a narrow, blue-shifted component that indicates that CO is released from the source primarily on the sun-facing side.  Given that all small bodies are spinning, with characteristic periods typically of order $t_s \sim$ 10 hr, 
the CO line asymmetry requires that the CO be very close to the physical surface.  To see this, consider a material in which the thermal diffusivity is $\kappa$ = 10$^{-7}$ m$^2$ s$^{-1}$, as representative of a finely powdered dielectric.  The diurnal thermal wave has a characteristic scale $\ell \sim (\kappa t_s)^{1/2}$.  Substituting, we estimate $\ell \sim$ 0.06 m (6 cm).  If the CO is buried any deeper than $\ell$, the heating of the CO will be unable to respond to the day-night insolation cycle, and the CO line profile will not show the sunward peak that is observed.  This argument seems robust in 29P and proves that the CO is very close to the surface of this nucleus.  However, we need more line profile measurements in other Centaurs in order to be sure that sunward outgassing is a general property of these objects.

\subsection{Amorphous Ice in the Centaurs}
\label{sectionamorphous}

Amorphous water ice is thermodynamically unstable and converts exothermically to the crystalline form at a rate that depends on the 
ice temperature.  The energy released 
per unit mass of pure water ice is about 9$\times$10$^4$ J kg$^{-1}$, and the timescale for the conversion is given in years by 

\begin{equation}
\tau_{cr} = 3.0\times10^{-21} \exp \left[\frac{E_A}{kT}\right]
\label{crystal}
\end{equation}

\noindent where $E_A$ is the activation energy, $k$ is Boltzmann's Constant and $E_A$/$k$ = 5370 K (Schmitt et al.\ 1989).  
Setting $\tau_{cr}$ = 1 $hr$ ($\sim$10$^{-4}$ yr), to correspond to the timescale of a typical laboratory experiment, the crystallization temperature
is given by Equation (\ref{crystal}) as $T_{cr}$ = 140 K.  In Solar system bodies where the timescales are much longer, significant crystallization can occur at correspondingly lower ice temperatures.  At temperatures $T <$ 77 K, the crystallization time exceeds the age of the Solar system.  Amorphous ice is highly porous and can contain within its structure a large number of atoms and molecules trapped at the time of condensation of the ice from vapor.  Experiments show that the trapped component is released upon crystallization, as the amorphous structure rearranges into a crystalline lattice and the trapped molecules are squeezed out (Notesco et al.\ 1996; 2003).  On laboratory timescales this happens
at $\sim$140 K but in Solar system bodies the gas release would occur more slowly at lower temperatures.   The timescale given in Equation (\ref{crystal}) is subject to uncertainties of extrapolation (the timescale is measured at temperatures much higher than relevant in the outer Solar system) and might be very different for ice formed in the presence of contaminants or particles, like silicate dust, that could act as crystallization nuclei.

The possible role of crystallization in the activity of comets has been studied in great detail (e.g.\ Enzian et al.\ 1997; Klinger et al.1996; Notesco et al.\ 1996; 2003; Prialnik 1997).  By providing an extra energy source, crystallization enables these models to be broadly successful in explaining
mass-loss from comets at temperatures too low (and heliocentric distances too large) for crystalline water ice to sublimate.  The crystallization models can be criticized on several levels.  Disconcertingly, there is no direct evidence for the existence of amorphous ice in cometary nuclei (however, in fairness we note that observations provided no direct evidence for water ice in the comets for many decades after the formulation of Whipple's (1950) dirty ice nucleus model).  The results of crystallization models depend upon a large number of unknown or poorly constrained parameters (e.g.\ thermal diffusivity, porosity, tortuosity, bulk composition, optical parameters including albedo and emissivity, and rotational properties) for each of which values must be assumed in order to fit the data.  Kouchi and Sirono (2001) have further argued that the transition in real ice might not be exothermic at all, because the sublimation of trapped volatiles liberated at crystallization would carry away all of the energy released.  We feel that this objection has merit within a mean free path of the escape surface but, in buried ice, the volatiles cannot escape directly to space and the energy of the phase transition would remain trapped.   


While observational evidence for amorphous ice in the Centaurs is, at best, indirect, evidence for  water ice clathrate is entirely lacking.  The clathrates trap guest molecules within the periodic lattice structure of crystalline water ice.  Hence, the 
volatility of clathrate ice is essentially the same as that of unclathrated, crystalline water ice (Klinger et al.\ 1996).  The perihelion distribution of the active Centaurs, which extends far beyond the range of distances over which crystalline water ice can sublimate (Figure \ref{ae}) is therefore inconsistent with the sublimation of clathrated water ice.  Furthermore, the conditions of low
temperature and low pressure likely to be found in the protoplanetary disk of the Sun, where the Centaurs formed, are more conducive to the
formation of amorphous forms than they are to clathrates (Devlin 2001).  If clathrates exist in the Centaurs, they have nothing to do with the observed outgassing activity.

Here, we suggest a simple estimate of where in the Solar system, if exposed amorphous ice is present, the phase transition may be important by comparing the crystallization time, 
$\tau_{cr}$, to the orbital period, $t_k$.  We reason that this gives a highly conservative criterion, because we can be sure that amorphous ice will have crystallized on objects for which $\tau_{cr} \ll t_k$ but crystallization may also have occurred on a longer timescale, up to a maximum set by the evolution of the orbit of the Centaur.   In the spirit of the criterion, we consider circular orbits, for which  $t_k = R_{au}^{3/2}$ [yr].  

We bracket the range of likely surface temperatures as in Section (5) using Equation \ref{sublimation} 
with $dm$/$dt$ = 0.  The highest surface temperature should be that at the subsolar point, for which $\chi$ = 1, giving
$T(R_{au})$ = $T_1 / R_{au}^{1/2}$ with $T_1$ = 392 K.  The lowest surface temperature is attained using the
isothermal blackbody approximation, for which  $\chi$ = 4 and
$T_1$ = 280 K.  Then, from Equation (\ref{crystal}) the
critical distance inside which exposed ice must be crystalline is found from the solution to 

\begin{equation}
\frac{3}{2} \ln R_{au} = -47.26 + \frac{E_A \sqrt{R_{au}}}{kT_1}
\label{critical}
\end{equation}
  
\begin{figure}[]
\begin{center}
\includegraphics[width=0.85\textwidth]{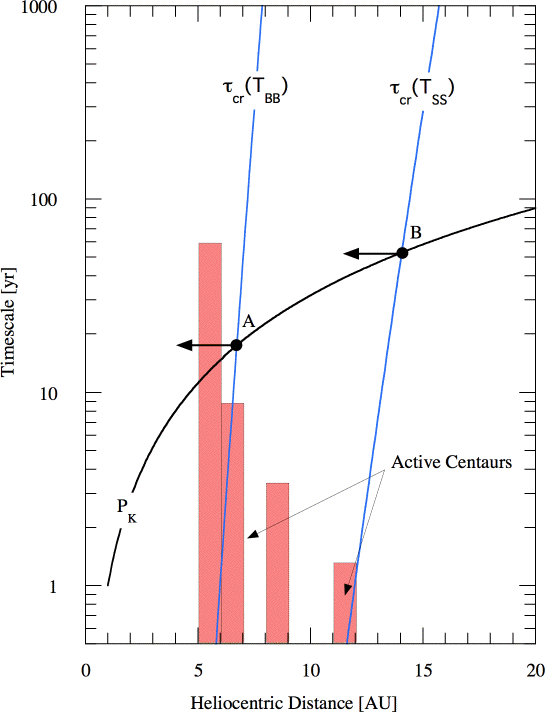}
\caption{The black line marked $P_K$ shows the variation of the Keplerian period
with semimajor axis.  Blue lines show the crystallization timescales for the lowest and highest
likely equilibrium temperatures, as described in Section \ref{sectionamorphous}.   The crystallization and
orbital periods are equal at the points labeled $A$ and $B$, having heliocentric distances
of $\sim$6.5 AU and 14 AU, respectively, computed from Equation (\ref{critical}).  Left-pointing arrows indicate that crystallization is expected
at all smaller heliocentric distances.  The histogram shows the distribution of the perihelion
distances of the active Centaurs.   
\label{TvsR}   } 
\end{center} 
\end{figure}

Solving for 280 $< T_1 <$ 392 K gives 6.8  $< R_{au} <$ 14.0 for  the distances inside which the phase transition is expected (see also Figure \ref{TvsR}). 
All of the active Centaurs have perihelia small enough for the crystallization of amorphous water ice to be a contributing factor.  This observation is suggestive, albeit indirect, evidence for the existence of amorphous ice in the Centaurs.

%

There are obvious difficulties in translating the results of laboratory experiments conducted typically on
timescales of hours to the relevant cometary timescales, measured in millions of years and longer.  Still, the overall picture is very attractive.
The Centaur precursors in the Kuiper belt contain amorphous ice, laden with trapped molecules including CO.  Once deflected into the
planetary region, a thermal wave of growing amplitude begins to propagate beneath the surface.  Regions within the Centaurs that attain
peak temperatures above $T_{cr}$ begin to convert to the crystalline phase, releasing trapped gases and the energy of the transition.   This occurs preferentially on the small-perihelion Centaurs but post-perihelion activity can be expected because of the finite timescales for the transport of heat in porous media.

Most of the active Centaurs were discovered soon before perihelion and were active at discovery.  This is shown in Table \ref{activeonset},
where we list the date of perihelion, the date of the first reported activity, the lag between these two dates, $\Delta T$, the orbital period, $P$, and the ratio $\Delta T$/$P$.  The interval between these dates, $\Delta T$, is typically measured in months,
only a small fraction of the orbital period, also listed in the Table.  Discovery near perihelion is an artifact of observational bias in magnitude
limited surveys, and holds no significance for us here.  The observation that mass-loss is underway close to and even before perihelion is more interesting.
It tells us that there is in general no great thermal lag between the period of maximum heating and the on-set of mass-loss.  The thermally driven process behind the production of coma is located close to the surface, just as we concluded earlier from the blue-shifted CO emission lines observed in 29P.

Observational evidence relating to the physical state of ice on the Centaurs is limited.  (10199) Chariklo (formerly 1997 CU26) shows water absorptions at 1.5 $\mu$m and 2.0 $\mu$m but there is no evidence for the 1.65 $\mu$m feature that is characteristic of crystalline ice (Dotto et al.\ 2003).  Given the quality of the spectrum and the muted depths of the 1.5 $\mu$m and 2.0 $\mu$m features, though,  the absence of the 1.65 $\mu$m band cannot be used as evidence for amorphous ice.  (2060) 95P/Chiron likewise shows water ice absorptions (Foster et al.\ 1999, Luu et al.\ 2000) but with absorption band strengths that are modulated by the activity of the coma.  The bands are again too weak and the data are too poor to permit us to rule definitively on the crystalline vs. amorphous state of the ice.   The activity in 2060 Chiron has been modeled as a product of crystallization heating by Prialnik et al. (1995).

\subsection{Amorphous Ice in the Comets and Kuiper Belt Objects}

The available data are consistent with the presence of amorphous ice in the Centaurs.   What does this mean for the dynamically related populations of Jupiter family comets (JFCs) and Kuiper belt objects?    

A fraction of the Centaurs will eventually be captured as JFCs, suggesting that these bodies should also contain amorphous ice.  Observationally, direct measurements of the ice in comets are very few, and the crystalline state is not well measured.  However, as noted above, crystallization has long been invoked by modelers as an important process in the JFCs, with an apparent success that is tempered by knowledge of the large number of poorly constrained parameters that must be assumed.  Crystallization models also provide plausible fits to the activity of some long-period and dynamically new (Oort cloud) comets (Meech et al. 2009).  A fair statement would be that there is no direct evidence concerning the crystallinity of cometary water ice and nothing to strongly contradict the presence of ice in the amorphous state.

A somewhat different circumstance prevails for the precursors to the Centaurs located in the Kuiper belt.  There, near infrared spectral observations provide a seemingly contradictory result; the 1.65 $\mu$m absorption band which is diagnostic of the presence of crystalline water ice is clearly observed in many KBOs (Jewitt and Luu 2004, Takato et al.\ 2006, Trujillo et al.\ 2007).  Some fraction of the ice could be amorphous and still fit the measured profile of the 1.65 $\mu$m band (Merlin et al.\ 2007, Pinilla-Alonso et al.\ 2009).  On the other hand, there are no well-observed objects in which the characteristic 1.5 $\mu$m and 2.0 $\mu$m bands of water ice are observed but the 1.65 $\mu$m crystalline ice band is absent.  Therefore, where the data are good enough to make a diagnosis, the available spectra provide no evidence that amorphous is ubiquitous on the KBOs.  If the Centaurs contain amorphous ice, and if they recently escaped from long-term storage in the Kuiper belt, it is logical to conclude that ice in the KBOs should be, at least in part, amorphous.  Certainly, amorphous ice could survive indefinitely in the KBOs given the very low radiation equilibrium temperatures in the Kuiper belt.    How might this seeming contradiction be explained?  

One possibility is that this is a size-selection effect.  The KBOs bright enough for the diagnostic 1.65 $\mu$n band to be spectroscopically detected tend to be large ($\sim$1000 km) objects.  Heating from their gravitational binding energy (McKinnon 2002) or from trapped radioactive nuclei may have caused the ice in these large bodies to have crystallized, so explaining the spectroscopic data (Merk and Prialnik 2003).   Smaller bodies of size comparable to the Centaurs listed in Table \ref{deltam} (radii of a few $\times$10 km or less) could have escaped crystallization but are effectively too faint to observe spectroscopically in the infrared. A second possibility is that the KBOs and Centaurs are stratified, with crystalline ice at the visible surface and amorphous ice beneath.  For example, some process such as heating by micrometeorite bombardment could crystallize surface ice while leaving deeper ice in the amorphous state.  Evidently, these two possibilities are not mutually exclusive.

\subsection{The Future}
A number of exciting observational prospects exist for making progress in this subject.  First, the assertion that the active Centaurs are powered by the crystallization of amorphous ice is open to an observational test:  this mechanism could be refuted if future observations reveal outgassing activity in Centaurs having perihelion distances much larger than the critical range for crystallization, as given by Equation (\ref{critical}).  For example, the discovery of
 active Centaurs with perihelion $q >$ 20 AU would be difficult to reconcile with the crystallization of amorphous ice, unless the dynamics showed that they had been
very recently scattered outwards from an orbit with a smaller perihelion.    Second, deep rotational line spectra with ALMA should show whether CO outgassing drives the mass-loss in other active Centaurs, as is expected in the present hypothesis.  Rotational line spectra having high sensitivity will also allow us to measure the mass-loss rates with greater accuracy, instead of relying on imperfect estimates based on observations of the dust, as in this work.  Third, measurements of the CO line profiles will show whether or not the mass-loss is mainly sunward (and therefore close to the physical surface), as is observed in 29P/Schwassmann-Wachmann 1.

Future observations might also show why some Centaurs remain inactive even when their perihelia are small.  In Figure (\ref{ae}), more than half of the Centaurs with perihelia between the orbits of Jupiter and Saturn appear inactive, although all are, presumably, hot enough to satisfy the crystallization criterion in Equation (\ref{critical}).  This could be because the mass-loss is intermittent or ``bursty'', on timescales longer than the orbital period, as a result of instabilities in the crystallization front. Alternatively, these hot but inactive Centaurs might have been devolatilized in their near-surface regions as a result of past activity (i.e. they could have been in their current orbits for longer, as a group, than the active Centaurs).  Still another possibility is that some Centaurs are deficient in volatiles, or perhaps lacking in the amorphous ice whose exothermic transformation appears to provide a plausible gas and energy source for the active Centaurs.   Physical studies, careful long-term photometric monitoring and dynamical investigations of the Centaurs will help us to understand which of these possibilities might be correct.

\clearpage 

\section{Summary}

We present new observations of the active Centaurs, defined as Solar system bodies whose orbits have both perihelia and semimajor axes between the orbits 
of Jupiter (at 5.2 AU) and Neptune (30 AU) and which are not in 1:1 resonance with the giant planets. 

\begin{enumerate} 
\item Including observations from the present work and from the literature, about 13\% (12 of 92) Centaurs are measurably active, as shown by the presence of dust comae in optical images. 

\item The active Centaurs have statistically smaller perihelion distances (median 5.9 AU) than the inactive
Centaurs in our sample (median 8.7 AU) and smaller than the median perihelion distance of all known Centaurs (12.4 AU).  Simulations show that the difference is unlikely to be an artifact of observational selection.  Instead, the smaller distances of the active Centaurs appear to reflect the action of a thermally-driven trigger or activity source.

\item The surface temperatures of the active Centaurs are too low (and the inferred mass-loss rates are too large) for the sublimation of crystalline water ice to constitute a plausible source of the activity.

\item Conversely, although gaseous carbon monoxide (CO) has been reported in two Centaurs (29P and Chiron), CO ice is too volatile to explain the  measured perihelion distribution.  If present, near-surface CO ice would strongly sublimate all the way out to Neptune's orbit, whereas no active Centaurs are known so far from the Sun.  


\item The active Centaurs have orbital periods which are long compared to the timescale for the crystallization of water ice at their distance. No active Centaurs in which this inequality is reversed have yet been found.   This fact is consistent with (but does not prove) cystallization as the trigger, or driver of activity.

 
 \item If Centaurs contain amorphous ice, as suggested by the available data, then their precursor bodies in the Kuiper belt must likewise be at least partly amorphous and their progeny, the nuclei of Jupiter family comets, should also contain amorphous ice.      
 
 \item Most active Centaurs are slightly redder than the Sun at optical wavelengths, but they appear devoid of the ultrared matter found in the inactive Centaurs and Kuiper belt objects.  Physical interpretation of the colors, however, is stymied by coma contamination in the active Centaurs.
 \end{enumerate}

\acknowledgments

I thank the referee, Humberto Campins, as well as Pedro Lacerda and Jing Li for their comments on the manuscript.  Some of the data were obtained at the W.M. Keck Observatory, which is operated as a scientific partnership by the California Institute of Technology, the University of California and the National Aeronautics and Space Administration. The Observatory was made possible by the generous financial support of the W.M. Keck Foundation.  I thank John Dvorak for operating the UH telescope, Joel Aycock,  Heather Hershley and Terry Stickel for running the Keck telescope and I am grateful for financial support from NASA's Planetary Astronomy and Origins Programs.

\renewcommand{\theequation}{A-\arabic{equation}}
  \setcounter{equation}{0}  
  \section*{APPENDIX}  

The connection between the cross-section of the particles and their mass depends largely upon the size distribution of the particles,
but also on their density and scattering efficiency.  
For spherical particles of a single radius, $a$, the mass, $M_d$, and the cross-section, $C_d$ = $\pi a^2$, are (in the geometric optics limit) proportional:

\begin{equation}
M_d =  \frac{4}{3} \rho a C_d
\label{mc}
\end{equation}

\noindent where $\rho$ is the grain bulk density.  

This simple relation may be generalized for power-law size distributions
in which the number of particles with radius between $a$ and $a + da$ is $\Gamma a^{-q} da$, where $\Gamma$ and $q$ are constants of the
distribution and the relation holds between minimum and maximum particle radii $a_{-}$ and $a_{+}$, respectively.  
The combined cross-section of such a distribution is

\begin{equation}
C_d = \pi \Gamma  \int_{a_{-}}^{a_{+}}  a^{2-q} da
\label{crosssection}
\end{equation}

\noindent while the total mass is 

\begin{equation}
M_d = \frac{4}{3} \pi \Gamma \int_{a_{-}}^{a_{+}}   a^{3-q} da
\label{dustmass}
\end{equation}

Results obtained from
the study of active comets show that the dust does not exactly follow a power law size distribution, but that a reasonable approximation
is obtained by setting $q$ = $3.5$,  
$a_{-}$ = 0.1 $\mu$m and $a_{+}$ = 1 cm, (e.g.\ Gr\"un et al.\ 2001).    Comparing Equations (\ref{crosssection}) and (\ref{dustmass}) then
shows that the mass is given by 

\begin{equation}
M_d =  \frac{4}{3} \rho \left(a_{-} a_{+}\right)^{1/2} C_d
\label{mass}
\end{equation}

\noindent and this is the relation we employ here.  The size of the average scatterer in this case is $\overline a$ = $(a_- a_+)^{1/2} \sim$ 30$\mu$m, consistent with the
observation that the comae of the Centaurs are neutral or reddish compared to the Sun, not blue (as would be expected if $\overline a \ll \lambda$, where $\lambda \sim 0.5~\mu m$ is the wavelength of observation).  

Other relations can be trivially computed from Equations (\ref{crosssection})
and (\ref{dustmass}) and still more involved estimates can be obtained by including the wavelength- and size-dependent scattering efficiencies
of the particles.  However, given our basic ignorance of the properties of the dust in Centaurs, these extra steps hardly seem appropriate.  In fact,
uncertainties in the dust parameters set a fundamental limit to our ability to estimate dust mass production rates from optical data and the absolute values of the mass-loss rates derived from Equation (\ref{mass}) could be wrong by an order of magnitude or more. In this paper, we use Equation (\ref{mass}), 
with $(a_- a_+)^{1/2} =$ 30$\mu$m and $\rho$ = 1000 kg m$^{-3}$, simply as a useful metric with which to compare relative mass-loss rates amongst the Centaurs.

\clearpage

\begin{deluxetable}{llrrrc}
\tabletypesize{\scriptsize}
\tablecaption{Journal of Observations \label{journal}}
\tablewidth{0pt}
\tablehead{
\colhead{Object} & UT Date & Tel &  \colhead{$R_{au}$\tablenotemark{a}}   & \colhead{$\Delta_{au}$\tablenotemark{b}}   &
\colhead{$\alpha$ [deg]\tablenotemark{c}}   }
\startdata

C/2001 M10 & 2002 Sep 08  & UH 2.2-m & 6.056 & 5.349  & 4.16 \\
C/2001 M10 & 2004 Oct 10  & Keck 10-m & 9.147 & 8.291 & 3.39  \\
P/2004 A1(LONEOS) & 2006 Aug 01 &	UH 2.2-m	& 6.079 &	5.961 & 9.62 \\
39P/Oterma & 2002 Sep 07  & UH 2.2-m  & 5.484 & 4.684 & 6.95  \\
29P/SW1    &  2002 Sep 08   & UH 2.2-m & 5.810 & 4.991& 6.24  \\
174P/Echeclus (60558) & 2006 Feb 25 & Keck 10-m & 12.968 & 12.230 & 3.00 \\
P/2005 T3 (Read) & 2007 Sep 19 & UH 2.2-m & 6.442 & 	6.368 &	8.97   \\
P/2005 T3 (Read) & 2007 Oct 12 & Keck 10-m & 6.459 & 	6.026 & 8.27   \\
P/2005 S2 (Skiff) & 2007 Oct 12 & Keck 10-m & 6.552 & 5.866 & 6.70 \\
165P/Linear  & 2003 Jan 08 & Keck 10-m & 8.300 & 8.123 & 6.75 \\
(63252) 2001 BL41 & 2002 Feb 22 & UH 2.2-m & 8.456 & 7.547 & 2.79 \\
(63252) 2001 BL41 & 2002 Feb 23 & UH 2.2-m & 8.458 & 7.555 & 2.89 \\
2000 GM137 & 2001 Feb 16 & UH 2.2-m & 6.970 & 6.348 & 6.62 \\
2000 GM137 & 2001 Feb 17 & UH 2.2-m & 6.970 & 6.336 & 6.54 \\
2003 QD112 & 2004 Oct 10 & Keck 10-m & 11.625 & 10.692 & 1.82 \\
(145486) 2005 UJ438 & 2007 Oct 12 & Keck 10-m &  9.280 & 8.609 & 4.73  \\
(32532) Thereus & 2007 Feb 20 & Keck 10-m & 10.792 & 11.233 & 4.60  \\
166P/2001 T4  & 2001 Nov 18  & UH 2.2-m & 8.590  & 7.841  & 4.49  \\
166P/2001 T4  & 2002 Sep 07  & UH 2.2-m & 8.573  & 7.937   & 5.43 \\
(5145) Pholus & 2004 Feb 17 & Keck 10-m & 18.538 & 18.488 & 3.05  \\
(148975) 2001 XA255 & 2002 Dec 07 & Keck 10-m & 15.062 & 14.262 & 2.24  \\
(148975) 2001 XA255 & 2003 Jan 09 & Keck 10-m & 14.965 & 13.982 & 0.14  \\
167P/2004 PY42 & 2004 Oct 10 & Keck 10-m & 12.234 & 11.840 & 4.36 \\
2006 SX368 & 2007 Sep 19 &UH 2.2-m & 12.398 & 11.703 & 3.44   \\
(54598) Bienor & 2006 Aug 23 & UH 2.2-m & 18.369 & 17.470 & 1.49  \\
2002 PN34 & 2006 Aug 23 & UH 2.2-m & 14.661  & 13.678 & 0.95  \\
2002 VG131 & 2002 Dec 07 & Keck 10-m & 14.870 & 14.431 & 3.24  \\
2001 XZ255 & 2003 Jan 29 & Keck 10-m & 16.079 & 15.096 & 0.09  \\
2001 XZ255 & 2004 Oct 10 & Keck 10-m & 16.167 & 16.385 & 3.43 \\
(52975) Cyllarus & 2004 Oct 10 & Keck 10-m & 21.089 & 20.542 & 2.30 \\

 \enddata

\tablenotetext{a}{Heliocentric distance in AU}
\tablenotetext{b}{Geocentric distance in AU}
\tablenotetext{c}{Phase angle in degrees}

\end{deluxetable}

\clearpage

\begin{deluxetable}{lrrcrr}
\tabletypesize{\scriptsize}
\tablecaption{Orbital Parameters of the Observed Centaurs\tablenotemark{a} \label{orbits}}
\tablewidth{0pt}
\tablehead{
 \colhead{Object} & $q$ [AU] \tablenotemark{b}& $a$\tablenotemark{c} &  \colhead{$e$\tablenotemark{d}}   & \colhead{$i$ [deg]\tablenotemark{e}}   &
\colhead{$Perihelion$}   }
\startdata

 C/2001 M10 & 5.303  & 26.660& 0.801& 28.0 & 2001 Jun 21 \\
 P/2004 A1(LONEOS) & 5.463 &	7.895	& 0.308 &	10.6 & 2004 Aug 25 \\
 39P/Oterma & 5.471  &7.256  & 0.246 & 1.9& 2002 Dec 21  \\
 29P/SW1    &  5.722   & 5.986 & 0.044 & 9.4& 2004 Jun 30  \\
 174P/Echeclus (60558) & 5.849 & 10.740 & 0.455 & 4.3 & 2015 May 06 \\
 P/2005 T3 (Read) & 6.202 & 	7.507 &	0.174	& 6.3 & 2006 Jan 13 \\
 P/2005 S2 (Skiff) & 6.398 & 7.964 & 0.197 & 3.1 & 2006 Jun 30 \\
 165P/Linear  & 6.830 & 18.03 & 0.621 & 15.9 & 2000 Jun 15 \\
 (63252) 2001 BL41 & 6.880 & 9.816 &  0.299 & 12.4 & 1998 Feb 09 \\
 2000 GM137 & 6.951 & 7.907 & 0.121 & 15.8 & 2002 Feb 24\\
 2003 QD112 & 7.935 & 18.974 & 0.582 & 14.5 & 1998 May 29 \\
 145486 & 8.255 & 17.527 & 0.529 & 3.8 & 2010 Jul 01 \\
 (32532) Thereus & 8.524 & 10.615 & 0.197 & 20.4 & 1999 Feb 15 \\
 166P/2001 T4  & 8.564   & 13.880& 0.383 & 15.4 & 2002 May 20  \\
 (5145) Pholus & 8.685 & 20.292 & 0.572 & 24.7 & 1991 Sep 24 \\
 (148975) 2001 XA255 & 9.387 & 30.183 & 0.689 & 12.7 & 2010 Jun 04 \\
 167P/2004 PY42 & 11.784 & 16.140 & 0.270 & 19.1 & 2001 Apr 10 \\
 2006 SX368 &    11.963 & 22.236 & 0.462 & 36.3 & 2010 May 16 \\
 (54598) Bienor &  13.155 & 16.485  & 0.202 & 20.8 & 2027 Dec 19 \\
 2002 PN34 &   13.328 & 30.781  & 0.567 & 16.7 & 2001 Nov 11 \\
 2002 VG131   & 14.869 &17.492 & 0.150 & 3.2 & 2002 Nov 09\\
 2001 XZ255  & 15.353 & 15.910 & 0.035 & 2.6 & 1984 Jun 22\\
 (52975) Cyllarus & 16.286 & 26.438 & 0.384 & 12.6 & 1989 Oct 27 \\

 \enddata

\tablenotetext{a}{Objects are listed in the order of increasing perihelion distance}
\tablenotetext{b}{Perihelion distance in AU}
\tablenotetext{c}{Semimajor axis in AU}
\tablenotetext{d}{Orbital eccentricity}
\tablenotetext{e}{Orbital inclination}

\end{deluxetable}

\clearpage

\begin{deluxetable}{llrrrrr}
\tabletypesize{\scriptsize}
\tablecaption{R-Band Photometry for Coma Search \label{deltam}}
\tablewidth{0pt}
\tablehead{
 \colhead{Object} & UT Date\tablenotemark{a}   & \colhead{$m_{2.2}$\tablenotemark{b}}   &  \colhead{$m_{3.3}$\tablenotemark{c}}  &
\colhead{$m_d$\tablenotemark{d}} &  \colhead{$m_R(1,1,0)$\tablenotemark{e}} &  \colhead{$r_e$ [km]\tablenotemark{f}}  }
\startdata

C/2001 M10 & 2002 Sep 08 & 18.85 & 18.53 & 20.02$\pm$0.04 & $>$10.92 & $<$12\\
P/2004 A1 & 2006 Aug 01 & 19.76 & 19.44 & 20.95$\pm$0.04 & $>$11.33 & $<$10\\
39P/Oterma & 2002 Sep 07 & 21.83 & 21.75 & 24.64$\pm$0.05 & $>$14.27 & $<$3 \\
29P/SW1 & 2002 Sep 08 & 16.32 & 15.86 & 17.01$\pm$0.05 & $>$8.52 & $<$35\\
174P/Echeclus (60558) & 2006 Feb 25 & 20.01 & 19.69 & 21.19$\pm$0.05 & $>$8.70 & $<$33 \\
P/2005 S2 (Skiff) & 2007 Oct 12 & 22.97 & 23.07 & $>$26.18 & 14.54 & 2\\
(63252) 2001 BL41 & 2002 Feb 22 & 20.35 & 20.33 & $>$23.85 & 11.03 & 11 \\
(63252) 2001 BL41 & 2002 Feb 23 &20.39 & 20.37 & $>$25.17 & 11.06 & 11\\
2000 GM137 & 2001 Feb 16 & 22.61 & 22.63 & $>$25.62 & 13.88 & 3\\
2000 GM137 & 2001 Feb 17 & 22.61 & 22.62 & $>$27.42 & 13.89 & 3 \\
2003 QD112 & 2004 Oct 10 & 22.32 & 22.01 & 23.55$\pm$0.04 & $>$11.62 & $<$9 \\
(145486) 2005 UJ438 & 2007 Oct 12  & 20.17 & 20.15 & $>$23.87 & 10.25 & 16\\
(32532) Thereus & 2007 Feb 20  & 19.87 & 19.87 & $>$23.39 & 9.05 & 28 \\
166P/2001 T4 & 2001 Nov 18 & 20.01 & 19.78 & 21.60$\pm$0.05 & $>$10.47 & $<$14 \\
166P/2001 T4 & 2002 Sep 07 & 19.65 & 19.27 & 20.63$\pm$0.05 & $>$10.04 & $<$18 \\
(5145) Pholus & 2004 Feb 17  & 19.96 & 19.94 & $>$25.51 & 7.00 & 72\\
(148975) 2001 XA255 & 2002 Dec 07  & 22.74 & 22.74 & $>$26.93 & 10.82 & 12\\
(148975) 2001 XA255 & 2003 Jan 09  & 22.39 & 22.30 & $>$26.20 & 10.74 & 13 \\
167P/2004 PY42 & 2004 Oct 10 & 20.69 & 20.64 & 24.04$\pm$0.05 & $>$9.50 & $<$23  \\
2006 SX368 & 2007 Sep 19  & 20.15 & 20.08 & 23.8$\pm$0.1 & 9.00 & 28\\
(54598) Bienor & 2006 Aug 23  & 21.11 & 21.53 & $>$25.90 & 8.38 & 38\\
2002 PN34 & 2006 Aug 23 & 19.92 & 19.89 & $>$26.23 & 8.26 & 40 \\
2002 VG131 & 2002 Dec 07  & 23.08 & 23.25 & $>$26.13 & 11.10 & 11\\
2001 XZ255 & 2003 Jan 09  & 22.44 & 22.34 & $>$26.14 & 10.48 & 14\\
2001 XZ255 & 2004 Oct 10 & 21.11 & 23.11 & $>$26.26 & 8.66 & 34 \\
(52975) Cyllarus & 2004 Oct 10 & 21.67& 21.66 & $>$26.29 & 8.23 & 41\\

 \enddata

\tablenotetext{a}{UT Date of the oberservation}
\tablenotetext{b}{Magnitude within a 2.2 arcsec radius aperture}
\tablenotetext{c}{Magnitude within a 3.3 arcsec radius aperture}
\tablenotetext{d}{Coma magnitude within the 2.2 to 3.3 arcsec annulus (Equation \ref{comamag})}
\tablenotetext{e}{Red magnitude of the nucleus, computed from $m_{2.2}$ corrected to unit heliocentric and geocentric distance and zero phase angle (Equation \ref{absolute})}
\tablenotetext{f}{Effective radius of the nucleus in km assuming red geometric albedo $p_R$ = 0.1.  Limits are given where coma was detected.}
\end{deluxetable}

\clearpage

\begin{deluxetable}{lrrrrrrrr}
\tabletypesize{\scriptsize}
\tablecaption{Model Mass Loss Rates \label{massloss}}
\tablewidth{0pt}
\tablehead{
\colhead{Object} & \colhead{$R$ [AU]\tablenotemark{a}} & $\Delta$ [AU]\tablenotemark{b}   & \colhead{$\alpha$ [deg]\tablenotemark{c}}   &  \colhead{$m_d$\tablenotemark{d}}  &
\colhead{$C_d$ [m$^2$]\tablenotemark{e}} &  \colhead{$\tau$ [s]\tablenotemark{f}} &  \colhead{$M_d$ [kg]\tablenotemark{g}} &  \colhead{$\frac{dM_d}{dt}$\tablenotemark{h}}   }
\startdata

C/2001 M10&	6.056&5.349&4.16&	20.0	&5.77$\times$10$^7$&8.16$\times$10$^3$&2.31$\times$10$^6$&2.8$\times$10$^2$ \\	
P/2004 A1	&6.079&5.961&9.62	&21.0&4.55$\times$10$^7$&9.10$\times$10$^3$&1.82$\times$10$^6$&2.0$\times$10$^2$ \\
39P/Oterma	&	5.484	&	4.684 & 6.95 & 24.6		&7.61$\times$10$^5$&6.97$\times$10$^3$&3.04$\times$10$^4$ & 4.3$\times$10$^0$ \\
29P/SW1 & 5.810 & 4.991 & 6.24 & 17.0 & 9.61$\times$10$^8$ & 7.53$\times$10$^3$ & 3.84$\times$10$^7$ & 5.1$\times$10$^3$ \\
174P & 12.970 & 12.230 & 3.00 & 21.2 & 1.85$\times$10$^8$ & 2.26$\times$10$^4$ & 7.41$\times$10$^6$ & 3.2$\times$10$^2$ \\
P/2005 S2 & 6.552 & 5.866 & 6.70 & $>$26.28 & $<$2.93$\times$10$^5$ & 9.12$\times$10$^3$ & $<$1.18$\times$10$^4$ & $<$1.3$\times$10$^0$\\
63252 & 8.458 & 7.555 & 2.89 & $>$25.17 & 	$<$1.23$\times$10$^6$ & 1.25$\times$10$^4$ & $<$4.93$\times$10$^4$ & $<$3.9$\times$10$^0$\\
2000 GM137 & 6.970 & 6.336 & 6.54 & $>$27.42 & 	$<$1.24$\times$10$^5$ & 1.00$\times$10$^4$ & $<$4.95$\times$10$^3$ & $<$0.5$\times$10$^{0}$\\
2003 QD112 & 11.625 & 10.692 & 1.82 & 23.5 & 1.33$\times$10$^7$ & 1.92$\times$10$^4$ & 5.31$\times$10$^5$ & 2.8$\times$10$^1$ \\
145486 & 9.280 & 8.609 & 4.73 & $>$23.87 & 	$<$6.63$\times$10$^6$ & 1.46$\times$10$^4$ & $<$2.65$\times$10$^5$ & $<$1.8$\times$10$^1$\\
32532 & 10.792 & 11.233 & 4.60 & $>$23.39 & $<$1.80$\times$10$^7$ & 1.98$\times$10$^4$ & $<$7.19$\times$10$^5$ & $<$3.6$\times$10$^1$\\
166P/2001 T4 & 8.573 & 7.937 & 5.43 & 20.6 & 1.10$\times$10$^8$ & 1.32$\times$10$^4$ & 4.40$\times$10$^6$ & 3.3$\times$10$^2$ \\
(5145) Pholus & 18.538 & 18.488 & 3.05 & $>$25.51 & $<$1.08$\times$10$^7$ & 3.73$\times$10$^4$ & $<$4.30$\times$10$^5$ & $<$1.2$\times$10$^1$\\
148975 & 15.062 & 14.262 & 2.24 & $>$26.93 & 	$<$1.37$\times$10$^6$ & 2.73$\times$10$^4$ & $<$5.49$\times$10$^4$ & $<$2.0$\times$10$^0$\\
167P/2004 PY42 & 12.234 & 11.840 & 4.36 & 24.0 & 1.31$\times$10$^7$ & 2.15$\times$10$^4$ & 5.24$\times$10$^5$ & 2.4$\times$10$^1$ \\
2006 SX368 & 12.398 & 11.703 & 3.44 & 23.8 & 1.52$\times$10$^7$ & 2.13$\times$10$^4$ & 6.10$\times$10$^5$ & 2.9$\times$10$^1$ \\
54598 & 18.369	&17.470 & 1.49 & $>$25.90 & $<$6.03$\times$10$^6$ & 3.52$\times$10$^4$ & $<$2.41$\times$10$^5$ & $<$6.9$\times$10$^0$\\
2002 PN34 & 14.661 & 13.678 & 0.95 & $>$26.23 &$<$2.11$\times$10$^6$ & 2.60$\times$10$^4$ & $<$8.45$\times$10$^4$ & $<$3.2$\times$10$^0$\\
2002 VG131 & 14.870 & 14.431 & 3.24 & $>$26.13 & $<$3.10$\times$10$^6$ & 2.75$\times$10$^4$ & $<$1.24$\times$10$^5$ & $<$4.5$\times$10$^0$\\
2001 XZ255 & 16.167 & 16.385 & 3.43 & $>$26.26 & 	$<$3.76$\times$10$^6$ & 3.19$\times$10$^4$ & $<$1.50$\times$10$^5$ & $<$4.7$\times$10$^0$\\
52975 & 21.089 & 20.542 & 2.30 & $>$26.29 & 	$<$7.03$\times$10$^6$ & 4.28$\times$10$^4$ & $<$2.81$\times$10$^5$ & $<$6.6$\times$10$^0$\\


 \enddata

\tablenotetext{a}{Heliocentric distance}
\tablenotetext{b}{Geocentric distance}
\tablenotetext{c}{Phase angle}
\tablenotetext{d}{Coma annulus magnitude from Table (\ref{deltam}).  We have taken the most sensitive limit for each object in which a coma was not detected}
\tablenotetext{e}{Derived dust cross-section (Equation \ref{inversesq})}
\tablenotetext{f}{Photometry annulus crossing time (Equation \ref{crossingtime})}
\tablenotetext{g}{Derived coma mass [kg], from $C_d$ and Equation \ref{mass}}
\tablenotetext{h}{Derived mass-loss rate [kg s$^{-1}$], from columns (g) and (f)}

\end{deluxetable}

\clearpage

\begin{deluxetable}{llccccc}
\tabletypesize{\scriptsize}
\tablecaption{Color and Multi-Aperture Photometry \label{newcolors}}
\tablewidth{0pt}
\tablehead{
\colhead{Object} & \colhead{UT 2005}    &  \colhead{$\phi$\tablenotemark{a}} & \colhead{R\tablenotemark{b}} & \colhead{B-V\tablenotemark{c}}  &  \colhead{V-R\tablenotemark{c}} &  \colhead{R-I\tablenotemark{c}} }
\startdata



29P/SW1  & 2002 Sep 08  &  1.1 & 16.56$\pm$0.01 &  0.79$\pm$0.03 & 0.51$\pm$0.03 & 0.52$\pm$0.03  \\
29P/SW1  & 2002 Sep 08  &  2.2 & 16.32$\pm$0.01 &  0.78$\pm$0.03 & 0.50$\pm$0.03 & 0.52$\pm$0.03  \\
29P/SW1  & 2002 Sep 08  &  3.3 & 15.86$\pm$0.01 &  0.79$\pm$0.04 & 0.47$\pm$0.03 & 0.50$\pm$0.03  \\
29P/SW1  & 2002 Sep 08  &  6.6 & 15.83$\pm$0.01 & 0.75$\pm$0.03 & 0.51$\pm$0.03 & 0.48$\pm$0.03 \\

39P/Oterma	& 	2002 Sep 07 	&   1.1 & 22.08$\pm$0.05 &  0.89$\pm$0.07 & 0.41$\pm$0.05 & 0.47$\pm$0.10  \\
39P/Oterma	& 	2002 Sep 07 	&   2.2 & 21.83$\pm$0.05 &  0.74$\pm$0.07 & 0.45$\pm$0.05 & 0.59$\pm$0.10  \\
39P/Oterma	& 	2002 Sep 07 	&   3.3 & 21.75$\pm$0.05 &  0.72$\pm$0.07 & 0.40$\pm$0.05 & 0.44$\pm$0.10  \\

166P/2001 T4 	& 	2001 Nov 18 &  1.1 & 20.47$\pm$0.03 &  0.87$\pm$0.04 & 0.69$\pm$0.03 & 0.78$\pm$0.04 \\
166P/2001 T4 	& 	2001 Nov 18 &  2.2 & 20.01$\pm$0.03 &   0.87$\pm$0.04 & 0.73$\pm$0.03 & 0.71$\pm$0.04 \\
166P/2001 T4 	& 	2001 Nov 18 &  3.3 & 19.78$\pm$0.03 &   0.87$\pm$0.04 & 0.66$\pm$0.04 & 0.73$\pm$0.04 \\
166P/2001 T4 	& 	2001 Nov 18 &  6.6 & 19.35$\pm$0.02 &  0.70$\pm$0.04 & 0.73$\pm$0.03 & -- \\

166P/2001 T4 	& 	2002 Sep 07	&   1.1   & 19.91$\pm$0.03  &  -- & 0.70$\pm$0.03 & 0.69$\pm$0.04 \\
166P/2001 T4 	& 	2002 Sep 07	&   2.2   & 19.65$\pm$0.03  &  -- & 0.63$\pm$0.03  & 0.70$\pm$0.04 \\
166P/2001 T4 	& 	2002 Sep 07	&   3.3   & 19.27$\pm$0.03 &  -- & 0.52$\pm$0.03 & 0.76$\pm$0.04  \\
166P/2001 T4 	& 	2002 Sep 07	&   3.3   & 19.03$\pm$0.02 &  -- & 0.52$\pm$0.03 & 0.76$\pm$0.04  \\





167P/2004 PY42 &  	2004 Oct 10 &  1.1  & 20.80$\pm$0.03 &  0.77$\pm$0.03 & 0.55$\pm$0.03 & 0.54$\pm$0.02 \\
167P/2004 PY42 &  	2004 Oct 10 &  2.2 & 20.69$\pm$0.02 & 0.80$\pm$0.03 & 0.51$\pm$0.03 & 0.51$\pm$0.03 \\
167P/2004 PY42 &  	2004 Oct 10 &  3.3 & 20.64$\pm$0.04 &  0.76$\pm$0.03 & 0.49$\pm$0.04 & 0.51$\pm$0.04 \\

C/2001 M10  &  2002 Sep 08 &  1.1 & 19.14$\pm$0.02 &  1.22$\pm$0.08\tablenotemark{d} & -- & -- \\
C/2001 M10  &  2002 Sep 08 &  2.2 & 18.85$\pm$0.02 &  1.19$\pm$0.08\tablenotemark{d} & -- & -- \\
C/2001 M10  &  2002 Sep 08 &  3.3 & 18.53$\pm$0.03 &  1.17$\pm$0.08\tablenotemark{d} & -- & -- \\
C/2001 M10  &  2002 Sep 08 &  6.6 & 18.52$\pm$0.04 &  1.16$\pm$0.08\tablenotemark{d} & -- & -- \\

C/2001 M10  &  2004 Oct 10 & 1.1 & 22.76$\pm$0.06 &  0.77$\pm$0.06 & 0.35$\pm$0.06 & 0.52$\pm$0.08 \\
C/2001 M10  &  2004 Oct 10 &  2.2 & 22.08$\pm$0.06 &  0.66$\pm$0.10 & 0.49$\pm$0.10 & 0.42$\pm$0.08 \\

174P/Echeclus (60558)  & 2006 02 25 &  1.1 & 20.27$\pm$0.02 & -- & -- & -- \\
174P/Echeclus (60558) & 2006 02 25 &  2.2 & 20.01$\pm$0.02 &  -- & -- & -- \\
174P/Echeclus (60558) & 2006 02 25 &  3.3 & 19.69$\pm$0.02 &  -- & -- & -- \\
174P/Echeclus (60558) & 2006 02 25 &  6.6 & 18.87$\pm$0.02 &  -- & -- & -- \\

2003 QD112 & 2004 10 10 &   1.1 & 22.49$\pm$0.03 &  -- & -- & -- \\
2003 QD112 & 2004 10 10 &  2.2 & 22.32$\pm$0.03 &  -- & -- & -- \\
2003 QD112 & 2004 10 10 &  3.3 & 22.18$\pm$0.03 &  -- & -- & -- \\
2003 QD112 & 2004 10 10 &  6.6 & 22.01$\pm$0.05 &  -- & -- & -- \\


P/2004 A1 & 2006 08 01 &  1.1  &  20.35$\pm$0.02 &  -- & -- & -- \\
P/2004 A1 & 2006 08 01 & 2.2 & 19.76$\pm$0.02 &  -- & -- & -- \\
P/2004 A1 & 2006 08 01 & 3.3 & 19.44$\pm$0.02 &  -- & 0.32$\pm$0.03 & -- \\
P/2004 A1 & 2006 08 01 & 6.6 & 19.01$\pm$0.02 &  -- & 0.39$\pm$0.03 & --  \\


 \enddata

\tablenotetext{a}{Radius of the photometry aperture, in arcsec. }
\tablenotetext{b}{Apparent $R$-band magnitude and its 1$\sigma$ uncertainty. }
\tablenotetext{c}{Color indices within each aperture}
\tablenotetext{d}{The B-R color is reported on this object only}

\end{deluxetable}

\clearpage

\begin{deluxetable}{llccccc}
\tabletypesize{\scriptsize}
\tablecaption{Adopted Colors of the Active Centaurs \label{finalcolors}}
\tablewidth{0pt}
\tablehead{
\colhead{Object} & \colhead{UT 2005}    &  \colhead{$\phi$\tablenotemark{a}} & \colhead{R\tablenotemark{b}}  & \colhead{B-V\tablenotemark{c}}  &  \colhead{V-R\tablenotemark{c}} &  \colhead{R-I\tablenotemark{c}} }
\startdata

29P/SW1 (coma)  & 2002 Sep 08  &  2.2 & 16.32$\pm$0.02 &  0.78$\pm$0.03 & 0.50$\pm$0.03 & 0.52$\pm$0.03  \\

39P/Oterma (coma)	& 	2002 Sep 07 	&   2.2 & 21.83$\pm$0.05 &  0.74$\pm$0.07 & 0.45$\pm$0.05 & 0.59$\pm$0.10  \\

166P/2001 T4 (coma) & 2001 Nov 18 & 2.2 & 20.01$\pm$0.02 &  0.87$\pm$0.03 & 0.73$\pm$0.03 & 0.71$\pm$0.03 \\

(2060) 95P/Chiron\tablenotemark{d} & 2001 Jun 13 &  1.4 & 15.77$\pm$0.04 &  0.60$\pm$0.03 & 0.32$\pm$0.03 & 0.64$\pm$0.03 \\

174P/Echeclus (60558) \tablenotemark{e} & 2006 Mar 23 &  1.0 & 20.13$\pm$0.1 &  0.71$\pm$0.15 & 0.56$\pm$0.15 & -- \\
174P/Echeclus (60558)  \tablenotemark{f} & 2006 Feb 24 &  30 & 15.18$\pm$0.04 &  0.81$\pm$0.07 & 0.50$\pm$0.06 & 0.58$\pm$0.05 \\

C/2001 M10 (coma) & 2004 Oct 10 & 2.2 & 22.08$\pm$0.06 &  0.66$\pm$0.10 & 0.51$\pm$0.03 & 0.51$\pm$0.03 \\

167P/2004 PY42 (nucleus) 
&  2004 Oct 10 & 2.2 &  20.69$\pm$0.02&  0.80$\pm$0.03 & 0.49$\pm$0.10 &0.42$\pm$0.08 \\

P/2004 A1 & 2006 Oct 01& 3.3 & 19.76$\pm$0.02 &  -- & 0.32$\pm$0.03 & -- \\

  \textit{Solar Colors} &  & &   & \textit{0.67} & \textit{0.36} & \textit{0.35} \\

 \enddata

\tablenotetext{a}{Radius of the circular aperture within which the measurement was taken, in arcseconds}
\tablenotetext{b}{$R$-band magnitude }

\tablenotetext{c}{Color index, as listed, with 1$\sigma$ uncertainty}

\tablenotetext{d}{Romon-Martin et al.\ 2003}

\tablenotetext{e}{Rousselot 2008; this small aperture measurement samples the primary nucleus}


\tablenotetext{f}{Bauer et al.\ (2008); this large aperture measurement samples the primary and secondary nuclei plus the associated dust envelope}
\end{deluxetable}

\clearpage

\begin{deluxetable}{lllrrrl}
\tabletypesize{\scriptsize}
\tablecaption{Onset of Activity \label{activeonset}}
\tablewidth{0pt}
\tablehead{
\colhead{Object}   & \colhead{Perihelion Date\tablenotemark{a}}  &  \colhead{Activity Date\tablenotemark{b}} &  \colhead{$\Delta T$ [yr]\tablenotemark{c}} & P [yr]\tablenotemark{d} & $\frac{\Delta T}{P}$ &Comment  }
\startdata
C/2001 M10 & 2001 Jun 21 & 2001 Jun 20 & 0.0 & 137.7 & 0.00 & Active at discovery\\
P/2004A1(LONEOS)  & 2004 Aug 25 & 2004 Jan 13 & -0.6 & 22.2 &-0.03& Active at discovery\\
39P/Oterma &  2002 Dec 21 & 2002 Sep 07 & -0.3 & 19.5 & -0.02 & -- \\
29P/SW1    &   2004 Jun 30 & -- & -- & 14.6 &  -- & Always active \\
174P/Echeclus (60558)  & 2015 May 06 & 2006 Jan & -9.4 & 35.2 & -0.27 & -- \\
P/2005T3 (Read) &  2006 Jan 13 & 2005 Oct & -0.3 & 20.6 & -0.01 & Active at discovery \\
P/2005 S2 (Skiff) &  2006 Jun 30 & 2005 Oct & -0.8 & 22.5 & -0.04 & Active at discovery \\
165P/Linear  & 2000 Jun 15 & 2000 Jan 29 & -0.5 & 76.6 & -0.01 & Active at discovery \\ 
2003 QD112 & 1998 May 29 & 2004 Oct 10 & +4.4 & 82.7 & +0.05 & -- \\
(2060) 95P/Chiron  & 1996 Feb 14 & 1989 Apr 10 & -6.8 & 50.8 &  -0.13& Active since 1989\\
166P/2001 T4   & 2002 May 20  & 2001 Oct 15 & -0.6 & 51.9 & -0.01 & Active since discovery\\
167P/2004 PY42  & 2001 Apr 10 & 2004 Aug 11 & +3.3 & 64.8 & +0.05 & -- \\

 \enddata

\tablenotetext{a}{Date of closest perihelion }
\tablenotetext{b}{Dates of first reported cometary activity}
\tablenotetext{c}{Time lag, in years, between perihelion and first date of reported activity.  Negative means activity
was detected before perihelion}
\tablenotetext{d}{Orbital period, in years}

\end{deluxetable}

\clearpage

\clearpage

\clearpage

\clearpage

\clearpage

\clearpage

\clearpage

\clearpage

\clearpage

\clearpage

\clearpage

\clearpage



\clearpage

\clearpage

\clearpage


\end{document}